\documentclass[12pt]{article}
\pdfoutput=1

\usepackage[utf8]{inputenc}
\usepackage[left=2.55cm, right=2.55cm, top=2.55cm, bottom=2.55cm]{geometry}
\usepackage{amsmath,amssymb,amsbsy}
\usepackage{slashed}
\usepackage{xcolor}
\definecolor{lightblue}{rgb}{0.93, 0.95, 1.0} 
\usepackage{graphicx}
\usepackage{cancel}
\usepackage{url}
\usepackage{cancel}
\usepackage{tcolorbox} 
\usepackage{cite}
\usepackage[colorlinks=true,allcolors=darkpurple,pdfborder={0 0 0},linktocpage=false]{hyperref}
\usepackage{tabularx,booktabs}
\usepackage{multicol}
\usepackage{units}
\usepackage{xspace}
\usepackage[labelfont=bf]{caption}
\usepackage[section]{placeins}
\usepackage{enumitem}
\usepackage{ulem}
\usepackage{subcaption}
\usepackage{multirow}
\usepackage{comment}
\usepackage{verbatim}
\usepackage{ulem}
\useunder{\uline}{\ul}{}
\usepackage{array} 
\newcolumntype{C}[1]{>{\centering\arraybackslash}p{#1}}
\renewcommand{\arraystretch}{1.5}
\usepackage[compat=1.1.0]{tikz-feynman}
\definecolor{darkred}{rgb}{0.6,0,0}
\definecolor{darkpurple}{rgb}{0.5,0,0.5}

\def\abs[#1]{\lvert #1\rvert}
\def\L{\mathcal{L}}

\def\O{\mathcal{O}}
\def\hc{\text{h.c.}}
\def\BR{\text{BR}}

\def\RE{\, \text{Re}}

\def\z2{$\mathbb{Z}_2$}

\def\id{\mathbb{I}}

\def\U1L{$\mathrm{U(1)}_L$}

\definecolor{avblue}{rgb}{0.0, 0.0, 0.8}
\definecolor{asparagus}{rgb}{0.53, 0.66, 0.42}
\definecolor{aqua}{rgb}{0.4, 0.6, 0.7}

\newcommand {\ignore}[1]{}

\newcommand{\AddrHeidelberg}{%
  Max-Planck-Institut f\"{u}r Kernphysik, Saupfercheckweg 1, 69117 Heidelberg, Germany}

\newcommand{\AddrIFIC}{%
  Instituto de F\'{i}sica Corpuscular, CSIC-Universitat de Val\`{e}ncia, 46980 Paterna, Spain}

\newcommand{\AddrFISTEO}{%
  Departament de F\'{\i}sica Te\`{o}rica, Universitat de Val\`{e}ncia, 46100 Burjassot, Spain}

\begin{document}


\begin{center}
\vspace*{15mm}

\vspace{1cm}
{\Large \bf 
The Type-I Seesaw family
} \\
\vspace{1cm}

{\bf Salvador Centelles Chuli\'a$^{\text{a}}$, Antonio Herrero-Brocal$^{\text{b}}$, Avelino Vicente$^{\text{b,c}}$}

\vspace*{.5cm}
 $^{(\text{a})}$ \AddrHeidelberg \\\vspace*{.2cm} 
 $^{(\text{b})}$ \AddrIFIC \\\vspace*{.2cm} 
 $^{(\text{c})}$ \AddrFISTEO

\vspace*{.3cm}
\href{mailto:chulia@mpi-hd.mpg.de}{chulia@mpi-hd.mpg.de},
\href{mailto:antonio.herrero@ific.uv.es}{antonio.herrero@ific.uv.es},
\href{mailto:avelino.vicente@ific.uv.es}{avelino.vicente@ific.uv.es}
\end{center}

\vspace*{10mm}
\begin{abstract}\noindent\normalsize
We provide a comprehensive analysis of the Type-I Seesaw family of neutrino mass models, including the conventional type-I seesaw and its low-scale variants, namely the linear and inverse seesaws. We establish that all these models essentially correspond to a particular form of the type-I seesaw in the context of explicit lepton number violation. We then focus into the more interesting scenario of spontaneous lepton number violation, systematically categorizing all inequivalent minimal models. Furthermore, we identify and flesh out specific models that feature a rich majoron phenomenology and discuss some scenarios which, despite having heavy mediators and being invisible in processes such as $\mu \to e \gamma$, predict sizable rates for decays including the majoron in the final state.
\end{abstract}

\section{Introduction}
\label{sec:intro}

The discovery of neutrino oscillations~\cite{Kajita:2016cak,McDonald:2016ixn} highlights the need for Beyond Standard Model (BSM) physics. While neutrinos are massless within the Standard Model (SM) framework, their observed masses not only point towards a novel mechanism beyond the traditional Higgs mechanism but also require a robust explanation for their relatively small scale compared to the electroweak scale. Among various models proposed in the literature, the type-I seesaw mechanism~\cite{Minkowski:1977sc,Yanagida:1979as,Mohapatra:1979ia,GellMann:1980vs,Schechter:1980gr} stands out for its simplicity and elegance. In this framework, the smallness of neutrino mass is inversely related to a new, higher mass scale $M$, represented by the Majorana mass of newly introduced neutral fermions.

While elegant from a theoretical standpoint, the type-I seesaw model inherently implies that any phenomenological effects are suppressed by the small seesaw expansion parameter $\varepsilon^2 = \mathcal{O}\left(\Lambda_\text{EW}^2 / M^2 \right) = \mathcal{O}\left(m_\nu / M \right) \ll 1$, where $\Lambda_\text{EW} \sim 100$ GeV is the electroweak scale. However, this constraint is relaxed in genuine low-scale variants of the model, such as the linear~\cite{Akhmedov:1995ip,Akhmedov:1995vm,Malinsky:2005bi} and inverse~\cite{Mohapatra:1986bd,GonzalezGarcia:1988rw} seesaws. In these models,
phenomenological effects are not neutrino-mass suppressed, potentially leading to discernible traces in charged Lepton Flavor Violation (cLFV)~\cite{Bernabeu:1987gr,Gonzalez-Garcia:1988okv,Abada:2014kba,Lindner:2016bgg,Hagedorn:2021ldq}, through direct production of mediators at colliders~\cite{Dittmar:1989yg, Gonzalez-Garcia:1990sbd, Atre:2009rg,Aguilar-Saavedra:2012dga,Das:2012ii,Deppisch:2013cya,Antusch:2015mia,Deppisch:2015qwa,Hirsch:2020klk,Cottin:2022nwp,Chauhan:2023faf,Batra:2023mds,Batra:2023ssq}, via non-standard neutrino propagation effects~\cite{Escrihuela:2015wra,Blennow:2016jkn} or other low-energy probes~\cite{Abada:2012mc,Abada:2013aba,Miranda:2021kre,Schwetz:2021thj,Schwetz:2021cuj,Tang:2021lyn,Arrington:2022pon,Capozzi:2023ltl,Soleti:2023hlr}. For a short review on low-scale neutrino mass models see for example~\cite{Boucenna:2014zba}.

The existence of Majorana masses for neutrinos inherently implies the violation of lepton number symmetry. This violation can occur either explicitly or spontaneously. When the symmetry is global, the latter leads to the presence of a Nambu-Goldstone boson, known as the majoron ($J$)~\cite{Chikashige:1980qk,Chikashige:1980ui,Schechter:1981cv,Gelmini:1980re,Aulakh:1982yn}. In this work we analyze the \textit{Type-I Seesaw family}~\footnote{From now on, we use capital letters for the family and lowercase for the specific model.}, composed by the standard type-I seesaw model and its many low-scale variants. We classify the members of the family and explore novel neutrino mass models that feature rich phenomenology. Cosmological imprints of the majoron such as $\Delta N_\text{eff}$ \cite{Baumann:2016wac,Planck:2018vyg} could provide a complementary approach to the low-scale seesaw signatures discussed above. Here, we will focus on models which feature interactions between the majoron and charged leptons as their main phenomenological signature, opening new avenues for detecting and studying the effects of lepton number violation. 

The paper is organized as follows. We start with a pedagogical introduction of the Type-I seesaw family and the explicit lepton number breaking scenario in Secs.~\ref{sec:general} and \ref{sec:explicit}. In Sec.~\ref{sec:spontaneous} we focus on the more interesting case of spontaneous symmetry breaking (SSB), where we aim at classifying and analyzing all inequivalent minimal models of the Type-I seesaw family. We also point out those models in which the majoron phenomenology is not neutrino-mass suppressed and flesh them out in Sec.~\ref{sec:pheno}. Finally, we conclude with a summary in Sec.~\ref{sec:summary}. Additional technical details are given in two Appendices.

\section{The Type-I Seesaw family}
\label{sec:general}

Let us start by defining the Type-I Seesaw family. A model belongs to the Type-I Seesaw family if its neutral fermion mass matrix can be written as
\begin{align}
\mathcal M =\begin{pmatrix}
    0 & M_D\\
    M_D^T & M_F
    \end{pmatrix} \, ,
\end{align}
in the basis $(\nu_i, F_j)$, where $\nu_i$ ($i=1,2,3$) are the usual 3 SM neutrinos while $F_j$ ($j=1,\dots,n_F$) are $n_F$ heavy BSM neutral fermions. $M_D$ is a general $3 \times n_F$ matrix and $M_F$ is an $n_F \times n_F$ symmetric matrix. Furthermore, we demand the following condition in order to consider the model as part of the Type-I Seesaw family: 
\par\bigskip\noindent
\centerline{\begin{minipage}{0.8\textwidth}
The hierarchy $\left(M_D \, M_F^{-1}\right)_{ij} \ll 1 \, \, \forall \, i, j$ is satisfied. This allows one to expand the relevant physical quantities in powers of $\varepsilon = \O\left (M_D \, M_F^{-1}\right)$. These are nothing but the seesaw limit and expansion, respectively.
\end{minipage}}
\par\bigskip
\noindent Which implies
\par\bigskip\noindent
\centerline{\begin{minipage}{0.8\textwidth}
A light Majorana mass term for the SM neutrinos, $M_\nu \ll \Lambda_\text{EW}$, is generated at tree level.
\end{minipage}}
\par\bigskip
\noindent Under this assumption we can compute a general formula for $M_\nu$. By rotating the fields into the mass eigenstates, the symmetric matrix $\mathcal M$ is brought into diagonal form by means of a Takagi decomposition as
\begin{equation}
U^T \mathcal M \, U = \widehat{\mathcal M} \, ,
\end{equation}
where $U$ is a unitary matrix and $\widehat{\mathcal M} = \text{diag}(m_1,m_2,\dots, m_{3+n_F})$ is the neutral fermion mass matrix in the mass basis. The matrix $U$ can be expressed as
\begin{equation} \label{eq:Ufactor}
    U= \begin{pmatrix}
    \sqrt{\id_3 - P P^\dagger} & P \\
    -P^\dagger & \sqrt{\id_{n_F} - P^\dagger P}
    \end{pmatrix} \, \begin{pmatrix}
    U_l & 0 \\
    0 & U_h
    \end{pmatrix}\equiv U_1 \, U_2 \, .
\end{equation}
Here, $U_l$, $U_h$ and $U_{1,2}$ are $3\times 3$, $n_F\times n_F$ and $(3+n_F) \times (3+n_F)$ unitary matrices, respectively, $P$ is a $3 \times n_F$ matrix and we denote a general $n \times n$ identity matrix as $\id_n$. This factorization of the unitary matrix $U$ allows one to easily identify the role played by each factor. $U_1$ brings the neutral fermion mass matrix into a block-diagonal form, while $U_2$ finally diagonalizes, independently, the light and heavy sectors of the matrix. For instance, the light sector is diagonalized as 
\begin{equation}
 U_l^T M_\nu \, U_l = \text{diag}(m_1,m_2, m_3) \, ,  
\end{equation}
with $m_1,m_2, m_3$ the active neutrino masses. Let us now focus on the block-diagonalization of the mass matrix. By expanding $P$ in powers of $\varepsilon$,
\begin{equation} \label{eq:Pmatrix}
    P=\sum_{i=1}^\infty P_i \, ,
\end{equation}
with $P_i \sim \varepsilon^{i}$, at leading order in $\varepsilon$ one finds
\begin{align}
  \sqrt{\id_3 - P P^\dagger}  &= \id_3 + \O\left(\varepsilon^2\right) \, , \\
  \sqrt{\id_{n_F} - P^\dagger P} &= \id_{n_F} + \O\left(\varepsilon^2\right) \, ,
  \end{align}
and
\begin{align}
  P = P_1 + \O\left(\varepsilon^2\right) = M_D^* \, \left(M_F^{-1}\right)^\dagger + \O\left(\varepsilon^2\right)   \, .
\end{align}
Using these results, we find at leading order in $\varepsilon$
\begin{align}
 U_1^T \mathcal M \, U_1 \approx \begin{pmatrix}
   - M_D \, M_F^{-1} \, M_D^T  & 0\\
    0 & M_F 
    \end{pmatrix}  \, ,
\end{align}
and, therefore, 
\begin{equation} \label{eq:Mnu}
   M_\nu = - M_D \, M_F^{-1} \, M_D^T + \O\left(\varepsilon^2\right) \, ,
\end{equation}
the well-known seesaw formula~\cite{Minkowski:1977sc,Yanagida:1979as,Mohapatra:1979ia,GellMann:1980vs,Schechter:1980gr}. We can now identify some known models belonging to this family based on the structure of the $M_F$ and $M_D$ matrices, namely, the type-I seesaw, the inverse seesaw and the linear seesaw. We will focus on the minimal realizations of the inverse and linear cases, which require two different neutral fermions, $N$ and $S$. For simplicity, we will consider the same number of generations for both of them, $n_N = n_S = n_F / 2$. In this case, the most general matrices are given by
\begin{align}
 M_D = \begin{pmatrix}
    m_D & m_L 
    \end{pmatrix} \, ,  && M_F= \begin{pmatrix}
     \mu_N & m_R \\
    m_R^T & \mu_S
    \end{pmatrix}  \, ,
\end{align}
where $m_D$, $m_L$ and $m_R$ are two $3 \times n_N$ and one $n_N \times n_N$ general matrices, respectively, and $\mu_S$ and $\mu_N$ two $n_N \times n_N$ symmetric matrices. In order to distinguish between models we must express $M_\nu$ in Eq.~\eqref{eq:Mnu} in terms of the blocks of $M_F$ and $M_D$. In order to compute $M_F^{-1}$, we will consider separately the cases $m_R \neq 0$ and $m_R = 0$ and assume that $m_R$ is invertible in the former case.~\footnote{This is usually assumed, for instance, in the inverse and linear seesaws. Similar results can be obtained by assuming $\mu_N$ or $\mu_S$ to be invertible.} One obtains 
\begin{equation} \label{eq:MFinvGeneral}
 M_F^{-1} = \left\{ \begin{array}{cl}
    \displaystyle \begin{pmatrix}
   - \left( m_R^T \right)^{-1} \, \mu_S & \id_{n_S} \\
    \id_{n_N} & -m_R^{-1} \, \mu_N 
    \end{pmatrix} \,
    \begin{pmatrix}
    m^{-1} & 0 \\
    0 & (m^T)^{-1}
    \end{pmatrix} &\, , \hspace{1cm}  \text{if $m_R \neq 0$} \, , \\
    & \\
    \displaystyle \begin{pmatrix}
         \mu_N^{-1} & 0 \\
         0 & \mu_S^{-1} 
    \end{pmatrix} &\, , \hspace{1cm} \text{if $m_R = 0$} \, .
    \end{array} \right.
\end{equation}
where we have defined $m = m_R - \mu_N \left(m_R^T \right)^{-1} \mu_S $. Thus, the light neutrino mass matrix is given by
%
%
\begin{equation} \label{eq:MnuGeneral}
   M_\nu = \left\{ \begin{array}{l}
    \displaystyle \left( m_D \left(m_R^T\right)^{-1} \mu_S - m_L  \right) \, m^{-1} \, m_D^T + \left(m_L m_R^{-1} \mu_N - m_D  \right) \, (m^T)^{-1} \, m_L^T + \O\left(\varepsilon^2\right), \\
    \text{if $m_R \neq 0$} \, , \\
   \\
    \displaystyle -m_D \, \mu_N^{-1} m_D^T - m_L \, \mu_S^{-1} m_L^T + \O\left(\varepsilon^2\right), \\
   \text{if $m_R = 0$} \, .
   \end{array} \right.
\end{equation}

Eq.~\eqref{eq:MnuGeneral} constitutes a general result, valid for any model of the Type-I Seesaw family. Note that there is a continuous equivalence between both cases in this equation, i.e. taking the limit $m_R \to 0$ in the case $m_R \neq 0$ yields the same result as the exact case $m_R=0$, as expected.

%
%
%

The different models in the Type-I Seesaw family correspond to different hierarchies among the blocks in the $M_D$ and $M_F$ matrices. In principle, each block is independent, resulting in a unique hierarchy for every pair of blocks. This gives rise to many hierarchies and models, some of which are very popular while others are less well-known. For instance, one finds the usual type-I seesaw whenever there is no hierarchy among the blocks in $M_F$. Alternatively, if $\mu_N,\mu_S \ll m_R$, Eq.~\eqref{eq:MnuGeneral} simplifies to
\begin{align}
    M_\nu = \, &  m_D \, \left( m_R^T \right)^{-1} \mu_S m_R^{-1} m_D^T + m_L \, m_R^{-1} \mu_N \left( m_R^T\right)^{-1} \, m_L^T \nonumber \\
    -& m_D \left( m_R^T \right)^{-1}  \, m_L^T \,  - m_L \, m_R^{-1} m_D^T + \O\left(\varepsilon^2\right) \, ,
\end{align}
which could either lead to an inverse (if $m_L \ll m_D$ and $\displaystyle \frac{\mu_S}{m_R} \gg \frac{m_L}{m_D}$) or to a linear seesaw (if $m_L \ll m_D$ and $\displaystyle \frac{\mu_S}{m_R} \ll \frac{m_L}{m_D}$). In general, many possibilities exist. In the usual case of one high-energy scale in $M_F$ and one or two low-energy scales (one in $M_F$ and, possibly, one in $M_D$), all scenarios are summarized in Table~\ref{tab:Hierarchies1}. We refer to Appendix~\ref{sec:app1} for a comprehensive discussion considering all possible hierarchies.

\begin{table}[t!]
\begin{center}
\begin{tabular}{|c|c|c|c|c|}
\hline
\multicolumn{2}{|c|}{Hierarchies} & $\mu_N \ll \mu_S \sim m_R $  & $\mu_S \ll \mu_N \sim m_R $ & $\mu_N, \mu_S \ll m_R $ \\ \hline
\multicolumn{2}{|c|}{$ m_L \sim \Lambda_{\rm EW}$ } &  type-I   &    type-I & type-I \\ \hline
$m_L \ll \Lambda_{\rm EW}$ & $\frac{m_L}{m_D} \gg \frac{\mu_S}{m_R}$ & type-I & linear & linear\\ \cline{2-5} 
$m_L \ll \Lambda_{\rm EW}$ & $\frac{m_L}{m_D} \ll \frac{\mu_S}{m_R}$ & type-I & inverse &  inverse \\ \hline
\end{tabular}
\end{center}
\caption{Classification of different models featuring one high-energy scale and one or two low-energy scales in the $M_D$ and $M_F$ matrices into three distinct neutrino mass generation mechanisms: type-I seesaw, inverse seesaw, and linear seesaw.}
\label{tab:Hierarchies1}
\end{table}

We have just seen that different internal hierarchies lead to different mass generation mechanisms in the context of the Type-I Seesaw family. However, one might wonder if these distinct mechanisms represent genuinely different models or if there is some kind of underlying model below them; i.e. if we can find one Lagrangian describing all these models. We will demonstrate that the latter is true in the case of the explicit breaking of lepton number symmetry, $U(1)_L$, but not when the breaking is spontaneous.

\section{Explicit lepton number violation}
\label{sec:explicit}

We start with a pedagogical Section showing that all models with
explicit lepton number violation can be seen as equivalent to a type-I
seesaw with specific numbers of fermion singlet generations and matrix
textures. Let us again consider $n_N$ generations of the $N$ and $S$ fermion singlets, with the Lagrangian
\begin{equation} \label{eq:Lag}
  -\mathcal{L} = y_N \, \bar{L} \tilde{H} N + y_S \, \bar{L} \tilde{H} S + m_R \, \bar{N}^c S + \frac{\mu_N}{2} \, \bar{N}^c N + \frac{\mu_S}{2} \, \bar{S}^c S + \hc \, .
\end{equation}
Here $\tilde H = i \sigma_2 H^\ast$, $y_N$ and $y_S$ are two $3 \times n_N$ Yukawa matrices, $m_R$ is an $n_N \times n_N$ matrix with dimensions of mass and $\mu_N$ and $\mu_S$ are two $n_N \times n_N$ symmetric matrices, both with dimensions of mass. The Lagrangian in Eq.~\eqref{eq:Lag} violates lepton number explicitly. It can be expanded as
\begin{align}
  -\mathcal{L} =& \, y_N \, \bar{L} \tilde{H} N + y_N^\dagger \, \bar{N} \tilde{H}^\dagger L + y_S \, \bar{L} \tilde{H} S + y_S^\dagger \, \bar{S} \tilde{H}^\dagger L + m_R \, \bar{N}^c S + m_R^\dagger \, \bar{S} N^c \nonumber \\
  &+ \frac{\mu_N}{2} \, \bar{N}^c N + \frac{\mu_N^*}{2} \, \bar{N} N^c + \frac{\mu_S}{2} \, \bar{S}^c S + \frac{\mu_S^*}{2} \, \bar{S} S^c \, , \label{eq:LagExp}
\end{align}
Let us now combine the $N$ and $S$ singlet fermions into the multiplet
$F$, defined by
\begin{equation}
  F = \left( \begin{array}{cc} N & S \end{array} \right) \, .
\end{equation}
%
%
%
%
%
Now we can make use of the identity $\bar{\psi}^c \chi^c = \bar{\chi}
\psi$, valid for any two fermions $\psi$ and $\chi$, to find
$\bar{S}^c N = \bar{N}^c S$. This allows us to write Eq.~\eqref{eq:LagExp} as
\begin{equation} \label{eq:LagFinal}
  -\mathcal{L} = \, Y \, \bar{L} \tilde{H} F + \frac{M_F}{2} \, \bar{F}^c F + \hc \, ,
\end{equation}
with the \textit{dictionary}
\begin{align}
  Y &= \left( \begin{array}{cc} y_N & y_S \end{array} \right) \, , \label{eq:dic1} \\
  M_F &= \left( \begin{array}{cc}
    \mu_N & m_R \\
    m_R^T & \mu_S \end{array} \right) \, . \label{eq:dic3}
\end{align}
The Lagrangian of Eq.~\eqref{eq:LagFinal} is that of a type-I seesaw
with $n_F = 2 \, n_N $ generations of $F$ singlets and the specific matrix textures given by Eqs.~\eqref{eq:dic1} and \eqref{eq:dic3}. This proves the equivalence of all models in the type-I family to a specific texture of the standard type-I seesaw. This is not surprising: once we allow for a source of explicit breaking of lepton number in the Lagrangian, the singlets $N$ and $S$ cannot be distinguished and the model becomes a type-I seesaw. As we will show in the next Section, this is no longer true if the breaking of lepton number is spontaneous.

\section{Spontaneous lepton number violation}
\label{sec:spontaneous}

We now turn towards the more interesting case of spontaneous lepton number violation. Our goal is to analyze the minimal realizations of the Type-I Seesaw family with spontaneously broken global $U(1)_L$. In defining \textit{minimal}, we consider models featuring the fields listed in Tab.~\ref{tab:Fcharges}. In addition to the SM doublets $H$ and $L$, with lepton numbers $q_H=0$ and $q_L=1$, respectively, we will allow for the presence of the new fermions $N$ and $S$, as well as a second scalar doublet $\chi$ and the scalar singlet $\sigma$. The number of generations of $N$ and $S$ will again be denoted by $n_N = n_S$ and left unspecified. The lepton number of $N$ will be fixed to $q_N=1$, so that the Yukawa term $\bar{L} \tilde{H} N$ is allowed and $N$ can be identified with the usual right-handed neutrino. We also impose that $\chi$ couples to $S$ through the Yukawa term $\bar{L} \tilde{\chi} S$, which fixes its lepton number to $q_\chi = q_S-1$, where $q_S$ is the lepton number of $S$. Finally, the lepton number of $\sigma$ will be left as a free charge, denoted by $q_\sigma$. All neutral scalars will be assumed to get non-zero vacuum expectation values (VEVs),
\begin{equation}
\langle H^0 \rangle = \frac{v_H}{\sqrt{2}} \, , \quad
\langle \chi \rangle = \frac{v_\chi}{\sqrt{2}} \, , \quad
\langle \sigma \rangle = \frac{v_\sigma}{\sqrt{2}} \, .
\end{equation}
Therefore, $H$ and $\chi$ will be responsible for the breaking of the electroweak symmetry, whereas $\chi$ and $\sigma$ will break $U(1)_L$ if they have non-vanishing lepton numbers. The breaking of lepton number implies the existence of a physical Goldstone boson, the majoron, $J$~\cite{Chikashige:1980qk,Chikashige:1980ui,Schechter:1981cv,Gelmini:1980re,Aulakh:1982yn}. Its presence has important consequences for the different energy scales in our setup. Typically, the hierarchy $v_\chi \ll v_\sigma$ is phenomenologically required, as otherwise the majoron would have a sizeable doublet component, allowing the decay $Z \rightarrow J J$, which is strongly constrained by LEP~\cite{ALEPH:2005ab}. Moreover, the additional hierarchy $v_\chi \ll v_H$ ensures that the real component of $H^0$ will be SM-like, as demanded by current LHC data. We note that this is a natural hierarchy, since $m_\nu \propto v_\chi$ in many models. Finally, we must impose $M_D \ll M_F$ to guarantee the validity of the seesaw approximation, as required for the model to belong to the Type-I Seesaw family. While we remain agnostic about the hierarchies among Yukawa couplings, we assume that these will not be stark enough to overcome the VEV hierarchies, i.e. the hierarchies in the mass matrix are driven by the VEVs.

\begin{table}[tb!]
\begin{center}
\renewcommand{\arraystretch}{1.35} %
\setlength{\tabcolsep}{5pt} 
\begin{tabular}{| c | c | c |}
\hline
Fields & $SU(2)_L \otimes U(1)_Y$ & $U(1)_L $ \\ 
\hline
$H$ & $ (\textbf{2}, \frac{1}{2}) $ & 0 \\ 
$\chi$ & $ (\textbf{2}, \frac{1}{2}) $ & $q_S -1$ \\ 
$\sigma$ & $(\textbf{1}, 0)$ & $q_\sigma$ \\
\hline
$L$ & $(\textbf{2}, -\frac{1}{2})$ & 1 \\
$N$ & $ (\textbf{1}, 0) $ & 1  \\
$S$ & $ (\textbf{1}, 0) $ & $q_S$  \\ \hline
\end{tabular}
\end{center}
\caption{Lepton number and gauge electroweak charges of the particles in the Type-I seesaw family. The lepton number of $N$ is fixed by the term $\bar{L} \tilde{H} N$, while the lepton number of $S$ is model-dependent and sequentially fixes the lepton number of $\chi$ through the Yukawa term $\bar{L} \tilde{\chi} S$.}
\label{tab:Fcharges}
\end{table}

Some additional comments are in order:
\begin{itemize}
\item Models that do not include all the fields in Tab.~\ref{tab:Fcharges}, but only a subset of them, will also be considered in our analysis. 
\item Models with $q_S=1$ allow for a significant simplification. First of all, $q_H=q_\chi=0$ and the second scalar doublet $\chi$ is actually redundant, since its role can be played by the usual Higgs doublet $H$. Moreover, $q_N=q_S$, which implies that the multiplet $F = \left( \begin{array}{cc} N & S \end{array} \right)$ transforms consistently under $U(1)_L$.
\item We will limit our analysis to models with a single majoron.

\item Relaxing our minimality requirements lead to additional options. For instance, many other variations based on the inverse seesaw mechanism were studied in~\cite{CentellesChulia:2020dfh}.
\end{itemize}

It proves convenient to classify the models in the Type-I Seesaw family according to the texture of their neutral fermion mass matrix $\mathcal{M}$. The different possibilities are given by:

\begin{tcolorbox}[colback=lightblue, title=Class 1]
\begin{align} 
    \begin{pmatrix}
    0 & m_D & m_L\\
    m_D^T & \mu_N & m_R \\
    m_L^T & m_R^T & \mu_S 
    \end{pmatrix} \,  &&
    \begin{pmatrix}
    0 & m_D & m_L\\
    m_D^T & 0 & m_R \\
    m_L^T & m_R^T & 0 
    \end{pmatrix} \,  &&
    \begin{pmatrix}
    0 & m_D & 0\\
    m_D^T & \mu_N & m_R \\
    0 & m_R^T & \mu_S 
    \end{pmatrix} \label{eq:class1}
\end{align} 
\end{tcolorbox}

\begin{tcolorbox}[colback=lightblue, title=Class 2]
\begin{align}
    \begin{pmatrix}
    0 & m_D & m_L\\
    m_D^T & 0 & m_R \\
    m_L^T & m_R^T & \mu_S 
    \end{pmatrix} \,  &&
    \begin{pmatrix}
    0 & m_D & 0\\
    m_D^T &  0& m_R \\
    0 & m_R^T & \mu_S \\
    \end{pmatrix}
\end{align}
\end{tcolorbox}

\begin{tcolorbox}[colback=lightblue, title=Class 3]
\begin{align}
    \begin{pmatrix}
    0 & m_D & m_L\\
    m_D^T & \mu_N & m_R \\
    m_L^T & m_R^T & 0 
    \end{pmatrix} 
\end{align}
\end{tcolorbox}

\begin{table}[h!]
\begin{center}
\renewcommand{\arraystretch}{1.35} 
\setlength{\tabcolsep}{5pt} 
\begin{tabular}{| c | c |  }
\hline
Lagrangian term &  Covariance under $U(1)_L $ \\ \hline
$\bar{N}^c N $ &  $2$ \\ 
$\bar{S}^c S$ &  $2 \, q_S$  \\
$\bar{N}^c S$ & $q_S+1$  \\ \hline
\end{tabular}
\end{center}
\caption{Covariance under $U(1)_L$ of the different mass terms in the gauge singlet sector that must be present or forbidden in a model leading to a certain mass matrix. Since we are imposing the minimality condition that only one $\sigma$ singlet exists, then all the quantities in the second column must be either $0$, and then the term is explicitly present in the Lagragian, or equal to $\pm q_\sigma$, and then it is spontaneously generated, or forbidden in any other case.}
\label{tab:SCharges}
\end{table}

This classification into different classes of matrices is not arbitrary. As will be shown below, the conditions allowing for the more complex matrices (left) can be relaxed to give the simpler models (right). The hierarchies among the mass terms will determine the type of model (type-I, inverse, linear or hybrid). Tab.~\ref{tab:SCharges} shows the covariance of the mass terms in the gauge singlet sector. These results are useful to extract the possible lepton number charges of $S$ and $\sigma$ that lead to each mass matrix. 

Let us now present all the different minimal realizations of the Type-I Family with SSB. For each realization, we will show the Lagrangian and the mass mechanism. We will also discuss the majoron resulting in each realization and comment on its 1-loop coupling to charged leptons. This coupling is a crucial feature to distinguish among models and was computed analytically in a generic scenario in~\cite{Herrero-Brocal:2023czw}. For the sake of completeness, we summarize the main results of this reference in Appendix~\ref{app:coupling}. To the best of our knowledge, among the models in the Type-I Seesaw family, this coupling is only known for some realizations of the conventional type-I seesaw (model $C1a$)~\cite{Chikashige:1980qk,Chikashige:1980ui,Schechter:1981cv,Gelmini:1980re,Aulakh:1982yn,Pilaftsis:1993af,Garcia-Cely:2017oco,Heeck:2019guh,Herrero-Brocal:2023czw} and the inverse seesaw (model $C1b\slashed{\chi}$)~\cite{Herrero-Brocal:2023czw}, while other models are studied here for the first time.~\footnote{Although not exactly the same model as $C3$, this coupling was obtained for a related version with explicit lepton number breaking in~\cite{deGiorgi:2023tvn}.} In particular, we will comment on the neutrino mass suppression (or lack thereof) of the majoron coupling to charged leptons. We will regard the majoron couplings to charged leptons as \textit{neutrino mass suppressed} when they are proportional to the same VEV ratios as those found in the neutrino mass formula. For instance, in the usual type-I seesaw with three generations of right-handed neutrinos, one finds that the coupling of the majoron to charged leptons scales as $g_{Jee}\propto m_D \, m_D^\dagger /v_\sigma$~\cite{Pilaftsis:1993af,Garcia-Cely:2017oco,Heeck:2019guh,Herrero-Brocal:2023czw}. This is neutrino mass suppressed since $m_D \, m_D^\dagger /v_\sigma \sim v_H^2/v_\sigma \sim m_\nu$.~\footnote{A neutrino mass suppressed majoron coupling may, in principle, be sizable. As pointed out in \cite{Heeck:2019guh}, it is possible to use the matrix structure of $m_D$ to suppress $m_\nu$ as $-m_D m^{-1}_R m^T_D$ while keeping $m_D \, m^\dagger_D/v_\sigma$ large. However, such a cancellation would require some fine-tuning, so we will ignore this fact.} Models with a majoron coupling to charged leptons that is not neutrino mass suppressed will be regarded as \textit{enhanced}. A summary of all the results that follow are given in Tab.~\ref{tab:models_summary}.

\begin{table}[ht]
\centering
\setlength{\tabcolsep}{13pt} 
\renewcommand{\arraystretch}{1.5} 
\newcolumntype{C}{>{\centering\arraybackslash}m{2cm}} 
\begin{tabular}{|C|C|C|C|C|}
\hline
\textbf{Viable Models} & \textbf{Charges ($\boldsymbol{S}$, $\boldsymbol{\chi}$, $\boldsymbol{\sigma}$)} & $\boldsymbol{J \ell \ell \propto m_\nu}$ & \textbf{Scalar term} & \textbf{Type}\\
\hline
$C1a$ & $\left(1, 0, -2\right)$& Yes & $\emptyset$ & Type-I  \\
$C1b$ &  $\left(-1, -2, -2\right)$ & Yes & $\chi^\dagger H \sigma$ & Hybrid \\
$C1b\slashed{\chi}$ & $\left(-1, \emptyset, -2\right)$& Yes  & $\emptyset$ & Inverse \\
$C1b\slashed{\sigma}$ & $\left(-1, -2, -1\right)$& Yes & $\chi^\dagger H \sigma \sigma$ & Linear\\
$C2a$ & $\left(-\frac{1}{3}, -\frac{4}{3}, -\frac{2}{3}\right)$ & Yes & $\chi^\dagger H \sigma \sigma$ & Type-I \\
$C2a\slashed{\chi}$ & $\left(-\frac{1}{3}, \emptyset, -\frac{2}{3}\right)$& Yes & $\emptyset$& Type-I\\
$C2b$ & $\left(0, -1, -1\right)$ & No &$\chi^\dagger H \sigma$ & Hybrid\\
$C2b\slashed{\chi}$ & $\left(0, \emptyset, -1\right)$ & No & $\emptyset$& Inverse\\
$C3$ & $\left(-3, -4, -2\right)$ & No &$\chi^\dagger H \sigma \sigma$ & Linear \\
\hline
\hline
$C3\slashed{\chi}$ & $\left(-3, \emptyset, -2\right)$& \multicolumn{3}{|c|}{$m_\nu$ at 1-loop}\\
$C4$ & $\left(0, -1, -2\right)$ & \multicolumn{3}{|c|}{Minimal version is non-realistic}\\
\hline
\end{tabular}
\caption{Summary of all the possible minimal models in the Type-I Seesaw family. The fourth column `Scalar term' shows the term in the scalar potential that is not self-conjugate in the cases where it exists. In the models with the doublet $\chi$ this term is necessary to avoid a massless doublet majoron. For this reason, in the $C1b\slashed{\sigma}$ model, the scalar singlet $\sigma$ does not couple to the neutral fermions, but is needed to generate such a term.}
\label{tab:models_summary}
\end{table}

\subsection{Class 1}
\label{sec:FirstSet}

Let us start with Class 1 and consider the first matrix in Eq.~\eqref{eq:class1},
\begin{equation}
\label{eq:mat1}
 \begin{pmatrix}
    0 & m_D & m_L\\
    m_D^T & \mu_N& m_R \\
    m_L^T & m_R^T & \mu_S
    \end{pmatrix} \, .
\end{equation}
As already explained and discussed explicitly below, the other mass matrices (and hence models) in this class can be recovered from this one by just removing some of the fields in our general setup. The presence of the mass term $\bar{N}^c N$ forces $|q_\sigma| = 2$. In combination with the term $\bar{S}^c S$, this implies $q_S = \pm 1$. In this case, $\bar{N}^c S$ is generated, either from a $\sigma \bar{N}^c S$ Yukawa term (once $\sigma$ gets a VEV) or as a bare Lagrangian term. This leads to two inequivalent realizations, which we now discuss separately.
%
\subsubsection{Class 1a: $\boldsymbol{q_S = 1 \, \Rightarrow \, q_\chi = 0}$, $\boldsymbol{q_\sigma= -2}$}
\label{sec:C1a}

Since in this case $q_S = q_N$ and $q_H = q_\chi$, the Yukawa Lagrangian can be written as
\begin{align}
-\L =& \, \bar{L} \tilde{H} \left( y_N \, N  +y_S \, S \right) +\sigma  \left( \lambda  \, \bar{N}^c S + \frac{1}{2}\lambda_N \, \bar{N}^c N  + \frac{1}{2} \lambda_S \,\bar{S}^c S \right)+ \hc \nonumber \\
=& \, Y \, \bar{L} \tilde{H} F+ \frac{1}{2} \,  \sigma \, \bar{F}^c \, \Lambda \, F + \hc \, .
\end{align}
Here we have identified $\chi = H$, defined $Y$ as in Eq.~\eqref{eq:dic1} and introduced 
\begin{equation}
  \Lambda = \left( \begin{array}{cc}
    \lambda_N & \lambda \\
    \lambda^T & \lambda_S \end{array} \right) \, .
\end{equation}
It is clear that this scenario, which we denote as $C1a$, is a conventional type-I seesaw, as discussed in Sec.~\ref{sec:explicit}. The neutrino mass matrix will be given by $m_\nu \sim \frac{Y^2 v_H^2}{\Lambda v_\sigma} \sim \frac{\Lambda_\text{EW}^2}{\Lambda_{\rm H}}$, with $v_\sigma \sim \Lambda_{\rm H} \gg \Lambda_\text{EW}$ a large seesaw scale. The majoron phenomenology will therefore be neutrino mass suppressed in this model, as discussed extensively in the literature \cite{Chikashige:1980qk,Chikashige:1980ui,Schechter:1981cv,Gelmini:1980re,Aulakh:1982yn,Pilaftsis:1993af,Garcia-Cely:2017oco,Heeck:2019guh,Herrero-Brocal:2023czw}. 

\subsubsection{Class 1b: $\boldsymbol{q_S = -1 \, \Rightarrow \, q_\chi = -2}$, $\boldsymbol{q_\sigma = -2}$}
\label{sec:C1b}

With these charges, $m_R$ can exist as a bare mass and the Lagrangian is given by
\begin{equation}
-\L = y_N \, \bar{L} \tilde{H} N  +y_S \,  \bar{L} \tilde{\chi} S + m_R \bar{N}^c S + \frac{1}{2} \, \lambda_N \, \sigma\, \bar{N}^c N  + \frac{1}{2} \, \lambda_S \, \sigma^* \bar{S}^c S + \hc \, .
\end{equation}
Once the electroweak and lepton number symmetries are spontaneously broken we obtain,
\begin{align}
    m_D = \frac{v_H}{\sqrt{2}} \, y_N \, , \, \, m_L = \frac{v_\chi}{\sqrt{2}} \, y_S  \, , \, \, \mu_N = \frac{v_\sigma}{\sqrt{2}} \, \lambda_N \, , \,\,  \mu_S= \frac{v_\sigma}{\sqrt{2}} \, \lambda_S \, .
\end{align}
The VEV hierarchy $v_\chi \ll v_H$, required for the Higgs to be SM-like, implies
\begin{align}
    m_L \ll m_D \, , && \mu_N \sim \mu_S \, . 
\end{align}
To keep the doublet component of the majoron negligible, we must also impose the hierarchy $v_\chi \ll v_\sigma$. Depending on the hierarchies between $m_L$, $m_R$ and $\mu_{S, N}$ one finds either a type-I seesaw, a linear seesaw, an inverse seesaw or a hybrid scenario. In any case, this model ($C1b$) and the ones in this class will lead to a majoron phenomenology suppressed by neutrino masses.

We now consider the other two matrices in Class 1. By removing the singlet $\sigma$ we find the second matrix
\begin{align}
-\L &= y_N \, \bar{L} \tilde{H} N  +y_S \,  \bar{L} \tilde{\chi} S + m_R \bar{N}^c S + \hc \,\, \rightarrow \,\,
 \begin{pmatrix}
    0 & m_D & m_L\\
   m_D^T& 0 & m_R \\
    m_L^T & m_R^T & 0
    \end{pmatrix} \, .
    \end{align}
This model, as it is, features a purely doublet massless majoron and is thus not realistic. However, this can be solved by re-adding the scalar singlet $\sigma$ with charge $q_\sigma = -1$. This allows for the term $\chi^\dagger H \sigma \sigma$, which would lead to singlet-doublet mixing after symmetry breaking, hence suppressing the doublet component of the majoron. This would be the most straightforward version of the classic linear seesaw with SSB and was studied extensively in \cite{Fontes:2019uld}. We denote this model as $C1b\slashed{\sigma}$.
If we instead keep the singlet scalar $\sigma$ and remove the scalar doublet $\chi$ we obtain the model $C1b\slashed{\chi}$, which is the spontaneous version of the inverse seesaw \cite{GonzalezGarcia:1988rw}. In this case one recovers the third mass matrix in Class 1,
\begin{align}
    -\L &= y_N \, \bar{L} \tilde{H} N + m_R \bar{N}^c S +\frac{1}{2} \, \lambda_N \,  \sigma\, \bar{N}^c N  +\frac{1}{2} \, \lambda_S \, \sigma^* \bar{S}^c S + \hc \,\, \rightarrow  \,\, \begin{pmatrix}
    0 & m_D & 0\\
   m_D^T& \mu_N & m_R \\
    0 & m_R^T & \mu_S
    \end{pmatrix} \, .
\end{align}
%
\subsection{Class 2}
\label{sec:SecondSet}

We start with the mass matrix
 \begin{align}
   \begin{pmatrix}
    0 & m_D & m_L\\
    m_D^T & 0 & m_R \\
    m_L^T & m_R^T & \mu_S
 \end{pmatrix} \, ,
 \end{align}
and again look at the $U(1)_L$ covariance of the gauge singlet terms in Tab.~\ref{tab:SCharges}. Since the term $\bar{N}^c N$ must be forbidden in this scenario, $q_\sigma \neq \pm 2$. Again, two inequivalent scenarios arise.

\subsubsection{Class 2a: $\boldsymbol{q_S = -\frac{1}{3} \, \Rightarrow \, q_\chi = -\frac{4}{3}}$, $\boldsymbol{q_\sigma = -\frac{2}{3}}$}
\label{sec:C2a}

In this case, the Yukawa Lagrangian can be written as
\begin{equation}
-\L = y_N \, \bar{L} \tilde{H} N  +y_S \,  \bar{L} \tilde{\chi} S + \lambda  \, \sigma  \bar{N}^c S +\frac{1}{2} \lambda_S  \sigma^* \, \bar{S}^c S   + \hc \, ,
\end{equation}
with
\begin{align}
    m_D = \frac{v_H}{\sqrt{2}} \, y_N \, , \, \, m_L = \frac{v_\chi}{\sqrt{2}} \, y_S  \, , \, \, m_R = \frac{v_\sigma}{\sqrt{2}} \, \lambda \, , \,\,  \mu_S= \frac{v_\sigma}{\sqrt{2}} \, \lambda_S \, .
\end{align}
Again, we must impose the VEV hierarchy $v_\chi \ll v_H \ll v_\sigma$, which leads to 
\begin{align}
    m_L \ll  m_D && m_R \sim \mu_S\, \gg \Lambda_\text{EW} . 
\end{align}
The contribution of $m_L$ is actually subdominant and this model would be a type-I seesaw, with $m_\nu \sim v_H^2 / v_\sigma$. As such, the same conclusion holds when removing the doublet, in which case we find model $C2a\slashed{\chi}$, with the second mass matrix
\begin{align}
-\L = y_N \, \bar{L} \tilde{H} N  + \lambda  \, \sigma  \bar{N}^c S +\frac{1}{2} \lambda_S  \sigma^* \, \bar{S}^c S   + \hc \,\, \rightarrow \,\,
 \begin{pmatrix}
    0 & m_D & 0\\
   m_D^T& 0 & m_R \\
    0 & m_R^T & \mu_S
    \end{pmatrix} \, .
\end{align}
Both options will result in a majoron phenomenology suppressed by neutrino masses.

\subsubsection{Class 2b: $\boldsymbol{q_S = 0 \, \Rightarrow \, q_\chi =-1}$, $\boldsymbol{q_\sigma = -1}$}
\label{sec:C2b}

With this charge assignment, the Yukawa Lagrangian is given by
\begin{equation}
-\L = y_N \, \bar{L} \tilde{H} N  +y_S \,  \bar{L} \tilde{\chi} S + \lambda  \, \sigma  \bar{N}^c S +\frac{1}{2} \mu_S \bar{S}^c S   + \hc \, .
\end{equation}
As usual, several mass parameters are generated after SSB,
\begin{align}
    m_D = \frac{v_H}{\sqrt{2}} \, y_N \, , \, \, m_L = \frac{v_\chi}{\sqrt{2}} \, y_S  \, , \, \, m_R = \frac{v_\sigma}{\sqrt{2}} \, \lambda  \, , 
\end{align}
and, due to the VEV hierarchy $v_\chi \ll v_H \ll v_\sigma$, one finds
\begin{align}
    m_L \ll m_D \ll m_R\, . 
\end{align}
Like in model $C1b$, described in Sec.~\ref{sec:C1b}, the hierarchy between $m_L$ and $\mu_S$ determines if the model is an inverse seesaw, a linear seesaw or a hybrid. In this case, removing the singlet scalar is not an option, as it would render some sterile states massless, but we can recover the second matrix by removing the doublet,
\begin{align}
-\L = y_N \, \bar{L} \tilde{H} N  + \lambda  \, \sigma  \bar{N}^c S +\frac{1}{2} \mu_S\, \bar{S}^c S   + \hc \,\, \rightarrow \,\,
 \begin{pmatrix}
    0 & m_D & 0\\
   m_D^T& 0 & m_R \\
    0 & m_R^T & \mu_S
    \end{pmatrix} \, .
\end{align}
The resulting model, denoted as $C2b\slashed{\chi}$, is a pure inverse seesaw and was analyzed in \cite{Boulebnane:2017fxw}. Both models have enhanced majoron phenomenology, as we will flesh out in Sec.~\ref{sec:pheno}.

\subsection{Class 3}
\label{sec:ThirdSet}

Finally, we reach Class 3. Again, we consider the mass matrix
 \begin{align}
   \begin{pmatrix}
    0 & m_D & m_L\\
    m_D^T & \mu_N & m_R \\
    m_L^T & m_R^T & 0 
 \end{pmatrix} \, .
 \end{align}
By looking at the covariance under $U(1)_L$ of the singlet sector (see Tab.~\ref{tab:SCharges}) we can determine the allowed lepton number charges of the states in the model. The term $\bar{S}^c S$ must be forbidden in this model, which implies that $q_S \neq 0$ and $|q_\sigma| \neq 2 q_S$. Thus, the only solution is given by
\begin{equation}
    q_S = -3 \, \Rightarrow \, q_\chi = -4 \, , \,\, q_\sigma = -2
\end{equation}
Hence, the Yukawa Lagrangian can be written as
\begin{equation}
-\L = y_N \, \bar{L} \tilde{H} N  +y_S \,  \bar{L} \tilde{\chi} S + \lambda  \, \sigma^*  \bar{N}^c S +\frac{1}{2} \lambda_N  \sigma \, \bar{N}^c N   + \hc \, .
\end{equation}
After symmetry breaking,
\begin{align}
    m_D = \frac{v_H}{\sqrt{2}} \, y_N \, , \, \, m_L = \frac{v_\chi}{\sqrt{2}} \, y_S  \, , \, \, m_R = \frac{v_\sigma}{\sqrt{2}} \, \lambda \, , \,\,  \mu_N = \frac{v_\sigma}{\sqrt{2}} \, \lambda_N \, .
\end{align}
Finally, the usual VEV hierarchy $v_\chi \ll v_H \ll v_\sigma$ implies in this case
\begin{align}
    m_L \ll  m_D  \ll  m_R \sim \mu_N \, ,
\end{align}
where once again the last condition comes from the requirement that the sterile states have to be heavy, in the seesaw spirit. This is a linear seesaw with enhanced majoron phenomenology, which will be fleshed out in Sec.~\ref{sec:pheno}. To the best of our knowledge, this model has not been discussed before. A similar model was analyzed in \cite{deGiorgi:2023tvn}, where the lepton number breaking is explicit and not spontaneous, leading to a massive majoron.

\subsection{Related models}
\label{sec:FourthSet}

In addition to the models described previously, we highlight two additional ones that, while not meeting our predefined criteria, remain both potentially viable and interesting. Removing the scalar doublet from the Class 3 mass matrix yields
\begin{equation}
\begin{pmatrix}
0 & m_D & 0\\
m_D^T & \mu_N & m_R \\
0 & m_R^T & 0
\end{pmatrix} . 
\end{equation}
This configuration results in neutrinos that are massless at the tree level. Nonetheless, neutrino masses emerge at the 1-loop level, and thus we do not consider it a part of the type-I seesaw family. This economical texture of the mass matrix was first studied in~\cite{Dev:2012sg}. It was also explored in~\cite{Boulebnane:2017fxw}, where it was referred to as the Extended Seesaw.

Another model, not covered in our analysis, emerges from assigning the lepton number charges as $q_S = 0 \, \Rightarrow \, q_\chi = -1$, $q_\sigma = -2$, which leads to the mass matrix
\begin{equation}
\begin{pmatrix}
0 & m_D & m_L\\
m_D^T & \mu_N & 0 \\
m_L^T & 0 & \mu_S
\end{pmatrix} . 
\end{equation}
This charge assignment introduces an accidental $U(1)_H \times U(1)_\chi \times U(1)_\sigma$ symmetry in the scalar potential, resulting in two massless Goldstone bosons, with one being of doublet nature. It is possible to implement a realistic version of this model by deviating from our minimality conditions and incorporating additional singlet scalars. However, it is again beyond the scope of our analysis.


\section{Phenomenology of selected models}
\label{sec:pheno}

\begin{figure}[tb!]
  \centering
  \begin{subfigure}{0.42\linewidth}
    \includegraphics[width=\linewidth]{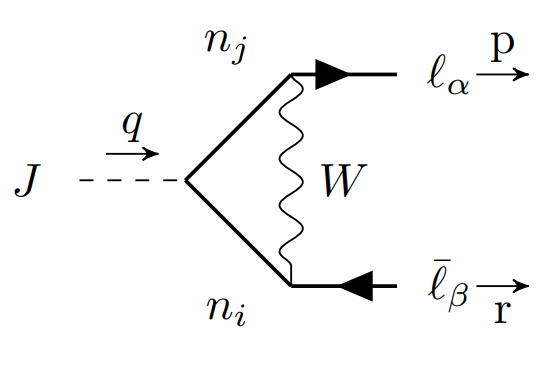}
    \caption{\textbf{$\boldsymbol{W}$ boson contribution}}
    \label{fig:Wdiag}
  \end{subfigure}
  \begin{subfigure}{0.57\linewidth}
    \includegraphics[width=\linewidth]{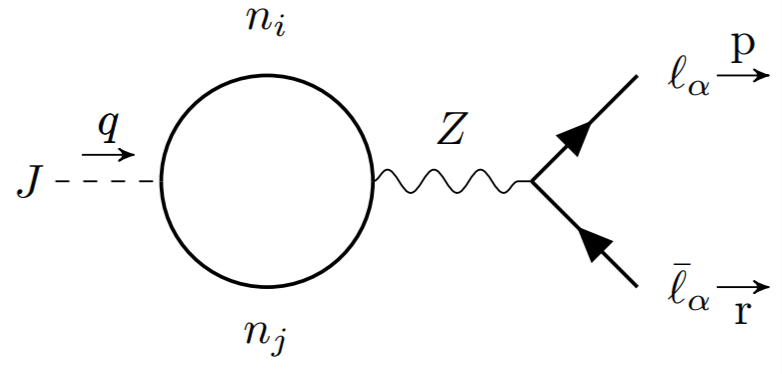}
    \caption{\textbf{$\boldsymbol{Z}$ boson contribution}}
    \label{fig:Zdiag}
  \end{subfigure}
  \vfill
  \bigskip
  \begin{subfigure}{0.42\linewidth}
    \includegraphics[width=\linewidth]{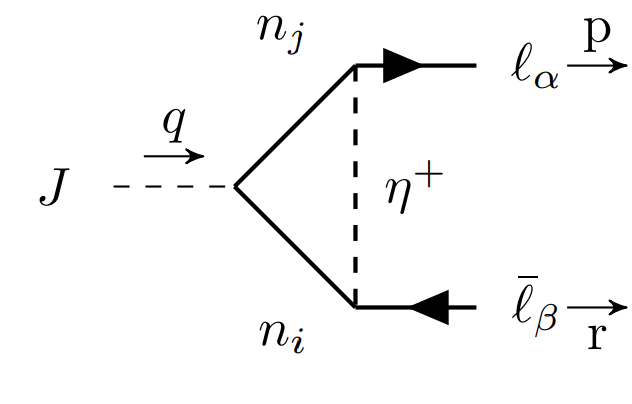}
    \caption{\textbf{$\boldsymbol{\eta}$ contribution}}
    \label{fig:EtaRhodiag}
  \end{subfigure}
  \begin{subfigure}{0.42\linewidth}
    \includegraphics[width=\linewidth]{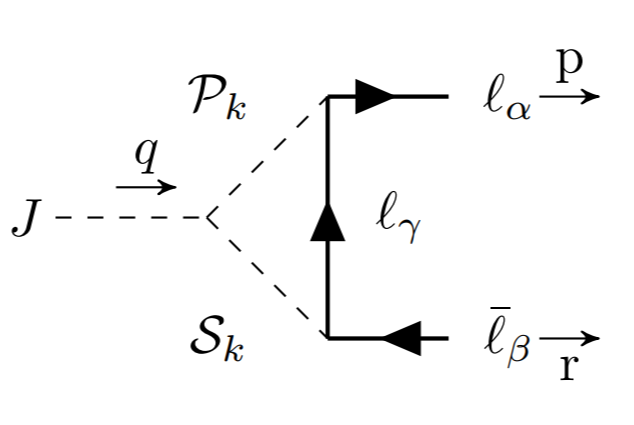}
    \caption{\textbf{Scalar contribution}}
    \label{fig:Sigmadiag}
  \end{subfigure}
  \vfill
  \bigskip
  \caption{Feynman diagrams leading to the 1-loop coupling of the
    majoron to a pair of charged leptons. \label{fig:diagrams}}
\end{figure}

The classification in the previous Section shows that the Type-I Seesaw family is composed by many distinct models, with potentially distinct phenomenological predictions. In particular, the relevance of the majoron coupling to charged leptons, a crucial feature that can be used to distinguish among models, has been highlighted. Reference~\cite{Herrero-Brocal:2023czw} recently computed general analytical expressions for the 1-loop coupling of the majoron to a pair of charged leptons, valid in any model with a clear hierarchy of energy scales, as required for the seesaw expansion to be consistent. In the context of the Type-I Seesaw family of models, one just has to compute the four diagrams shown in Fig.~\ref{fig:diagrams}.~\footnote{Other 1-loop diagrams represent corrections to a possible tree-level coupling and can be neglected.} In these diagrams, ${\cal S}_k$, ${\cal P}_k$ and $\eta^+$ represent, respectively, the neutral scalars, pseudoscalars and charged scalar in our setup. As already explained, full expressions of the resulting majoron coupling to charged leptons are given in~\cite{Herrero-Brocal:2023czw}. We collect the most relevant results of this reference and apply them to the Type-I Seesaw family in Appendix~\ref{app:coupling}.

We will now explore the majoron phenomenology of three specifically chosen models that feature an enhanced interaction of the majoron with charged leptons. As seen in Table~\ref{tab:models_summary}, the majoron interactions in the models not covered in this discussion are suppressed by neutrino masses.

\subsection{Hybrid mechanism with enhanced majoron LFV}
\label{subsec:Hybridenhanced}

\begin{table}[tb!]
\begin{center}
\renewcommand{\arraystretch}{1.35} 
\setlength{\tabcolsep}{5pt} 
\begin{tabular}{| c | c | c |}
\hline
Fields & $SU(2)_L \otimes U(1)_Y$ & $U(1)_L $ \\ 
\hline
$H$ & $ (\textbf{2}, \frac{1}{2}) $ & 0 \\ 
$\chi$ & $ (\textbf{2}, \frac{1}{2}) $ & $-1$ \\ 
$\sigma$ & $(\textbf{1}, 0)$ & $-1$ \\
\hline
$L$ & $(\textbf{2}, -\frac{1}{2})$ & 1 \\
$N$ & $ (\textbf{1}, 0) $ & 1  \\
$S$ & $ (\textbf{1}, 0) $ & $0$  \\ \hline
\end{tabular}
\end{center}
\caption{Lepton number and gauge electroweak charges of the particles in model $C2b$. This model leads to a hybrid inverse-linear seesaw mechanism for neutrino masses.}
\label{tab:C2b}
\end{table}

Let us start by analyzing model $C2b$. The lepton number charges of the fields in the model can be read off from Tables~\ref{tab:Fcharges} and \ref{tab:models_summary}, but we give them explicitly in Table~\ref{tab:C2b} for the sake of clarity. The model is defined by the Yukawa interactions given in Sec.~\ref{sec:SecondSet},
\begin{equation}
-\L = Y \bar{L} H e_R + y_N \, \bar{L} \tilde{H} N  +y_S \,  \bar{L} \tilde{\chi} S + \lambda  \, \sigma  \bar{N}^c S +\frac{1}{2} \, \mu_S\, \bar{S}^c S   + \hc \, ,
\end{equation}
which leads to the neutral fermion mass matrix
\begin{equation}
\mathcal{M} = \begin{pmatrix}
    0 & m_D & m_L\\
   m_D^T& 0 & m_R \\
    m_L^T & m_R^T & \mu_S
    \end{pmatrix} \, ,
\end{equation}
with $m_D = y_N \, v_H/\sqrt{2}$, $m_L = y_S \, v_\chi/\sqrt{2}$ and $m_R = \lambda \, v_\sigma/\sqrt{2}$. By assuming the usual hierarchies, $\mu_S, \, v_\chi \ll v_H\ll v_\sigma$, the neutrino mass matrix is found to be
\begin{equation}
\label{eq:mnuhybrid}
M_\nu \approx \frac{v_H^2}{v_\sigma^2} \left[y_N \, \left( \lambda^T \right)^{-1} \mu_S \lambda^{-1} y_N^T\right] \, - \frac{v_H \, v_\chi}{v_\sigma} \left[ \left( y_N \left( \lambda^T \right)^{-1}  \, y_S^T \right) \, + \text{tr.} \right] \, ,
\end{equation}
where tr. denotes the transposed of the previous matrix. The first term is the well-known neutrino mass matrix from an inverse seesaw, while the second is the one from a linear seesaw. Thus we deem this model as a hybrid mechanism. For the sake of generality we will be agnostic regarding the hierarchy $\mu_S / v_\chi$. $\lambda$ can be taken to be diagonal and real without loss of generality and, given the hierarchy $\mu_S \ll \lambda \, v_\sigma$,  $\lambda \, v_\sigma$ represents the physical masses of the quasi-Dirac pairs of heavy neutral leptons.

In the scalar potential, apart from the usual self-conjugate terms, the following terms are also present:
\begin{equation}
    V \supset  \lambda_{H\chi 2} \, (H \chi^\dagger) (\chi H^\dagger)+ \kappa  \, H \chi^\dagger \sigma + \hc \, .
\end{equation}
The CP-even scalar sector consists of three states: ${\cal S}_1 \equiv h$, ${\cal S}_2$ and ${\cal S}_3$. The lightest of them, $h$, is identified with the SM-like state discovered at the LHC with a mass $m_{{\cal S}_1} \approx 125$ GeV. Regarding the CP-odd scalars, there are two massless Goldstone bosons, ${\cal P}_1 \equiv G_Z$ (absorbed by the $Z$ boson) and ${\cal P}_2 \equiv J$ (the physical majoron), as well as the massive ${\cal P}_3$. Finally one of the two charged scalar states, $G_W$, is the usual EW Goldstone boson absorbed by the $W$ boson and there is a new massive charged state, $\eta^+$. Their masses are given as follows
\begin{itemize}
    \item CP-even scalars: $m_h^2\sim v_H^2$, $m_{{\cal S}_2}^2 \sim \kappa \, v_H \, \frac{v_\sigma}{v_\chi}$ and $m_{ {\cal S}_3}^2 \sim v_\sigma^2$
    \item CP-odd scalars: $m_{G_Z} = m_J = 0$ and $m_{{\cal P}_3}^2 = - \frac{\kappa \, v_\sigma v^2}{\sqrt{2} v_H v_\chi} \mathcal{N}^2$
    \item Charged scalars: $m_{G_W}=0$ and $m_{\eta^+}^2 = - \frac{\lambda_{H\chi 2}}{2} v^2 - \frac{\kappa}{\sqrt{2}} \,\frac{ v_\sigma v^2}{v_\chi v_H}$
\end{itemize}
where we have defined the normalization factor
\begin{equation} \label{eq:Nhybrid}
\mathcal{N}^2 = 1 + \frac{v_H^2 \, v_\chi^2}{v_\sigma^2 \, v^2} \, . 
\end{equation}
Here $v$ is the SM VEV, defined by $v^2 \equiv v_H^2 + v_\chi ^2 $. The mass-dimension trilinear $\kappa$ is a free parameter of the model with a large impact on the scalar spectrum. When $\kappa \gg \frac{v_\chi}{v_H} v_\sigma$, the BSM states ${\cal S}_2$, ${\cal P}_3$ and $\eta^+$ become heavy and, since we assume $v_\sigma$ to be a high-energy scale, decouple from the spectrum. Otherwise, our setup leads to a scalar spectrum with new states below $v_\sigma$, within the reach of current experiments. As expected, we find a SM-like Higgs and the Goldstone bosons associated to the SSB of the electroweak symmetry, in addition to the one from the SSB of $U(1)_L$. In the basis $( H^a, \chi^a, \sigma^a)$, where the $a$ superscript denotes the imaginary component of the corresponding field, the majoron $J$ is given as 
\begin{align}
     J &= \frac{1}{\mathcal{N}}\left(- \frac{v_H v_\chi^2}{v_\sigma v^2}, \,  \frac{v_H^2 v_\chi}{v_\sigma v^2}, \, 1 \right) \, , 
\end{align}
where $\mathcal{N}$ is the normalization factor of Eq.~\eqref{eq:Nhybrid}.
For the massive charged scalar, in the basis $(H^+, \, \chi^+)$, we find
\begin{align}
    \eta^+ &= \frac{v_H}{v}\left( -\frac{ v_\chi}{v_H}, \,  1 \right) \, .
\end{align}
We can now obtain the majoron coupling to charged leptons. First of all, note that the majoron has a component in the $H^a$ direction, so a tree-level coupling is generated, highly suppressed by the mixing:
\begin{align}
  \mathcal{L}_{\ell\ell J}^{\textup{tree-level}} = i \frac{v_\chi^2 v_H}{ \,\mathcal{N} \, v^3 v_\sigma} \, J \bar{\ell} M_\ell \, \gamma_5 \, \ell \sim \frac{v_\chi^2 \, m_\ell}{v^2 v_\sigma} \, J \bar{\ell} \, \gamma_5 \, \ell \lesssim \frac{v_\chi \, m_\ell \, m_\nu}{v^3} \, J \bar{\ell} \, \gamma_5 \, \ell \, .
\end{align}
However, we obtain larger couplings at 1-loop. To compute them we need the couplings for the dominant diagrams: $W$, $Z$ and $\eta^+$, i.e. diagrams (a), (b) and (c) of Fig.~\ref{fig:diagrams}, while diagram~(d) turns out to be subdominant. From Eqs.~\eqref{eq:Abar} and \eqref{eq:Dbar} we obtain, 
\begin{align}
\bar{A} &= \frac{1}{2} \begin{pmatrix}
0 & \bar{A}_L \\
\bar{A}_L^T & \bar{A}_H
\end{pmatrix} \, , \, \, \, \, \bar{A}_L = i\frac{v_\chi v_H}{\sqrt{2} \mathcal{N} v^2 v_\sigma}\begin{pmatrix}
-v_\chi y_N & v_H y_S 
\end{pmatrix} \, , \, \, \, \, \bar{A}_H = \frac{i}{\sqrt{2} \mathcal{N}} \begin{pmatrix}
    0 & \lambda \\
    \lambda^T & 0
\end{pmatrix} \, , \\
\bar{D}_R &= \frac{1}{v} \begin{pmatrix}
    - v_\chi \, y_N & v_H \, y_S
\end{pmatrix}   \, .
\end{align}
With this, it is trivial to obtain the necessary matrices. For the gauge boson diagrams we find,
\begin{align}
  \sum_{j \sim l } \Gamma_{\beta \alpha j}^{1,0,0} &= \sum_{j \sim l} \tilde{\Gamma}_{\beta \alpha j}^{1,0,0}\simeq - \frac{i}{2 v_\sigma}  \left (m_D m_D^\dagger \right)_{\beta \alpha} \, = - i\frac{v^2}{4 v_\sigma}  \left( y_N \, y_N^\dagger \right)_{\beta \alpha} \, , \\
   \sum_{j \sim h} \Delta_{\beta \alpha j}^{0,1,-1} &= \sum_{j \sim h} \tilde{\Delta}_{\beta \alpha j}^{0,1,-1} \simeq \frac{i}{ 2v_\sigma}  \left (m_D m_D^\dagger \right)_{\beta \alpha} = i\frac{v^2}{ 4 v_\sigma}  \left (y_N \, y_N^\dagger \right)_{\beta \alpha} \, ,
   \end{align}
while for the $\eta^+$ contribution, 
\begin{align}
  \sum_{j } \tilde{\Gamma}_{s p j}^{1,0,0} &\simeq \sum_{j \sim h} \Delta_{s p j}^{0,1,-1} \simeq \frac{i}{2 \mathcal{N} v_\sigma}\begin{pmatrix}
   m_R^2 & 0 \\
   0 & m_R^2 
  \end{pmatrix}_{sp} \simeq \frac{i v_\sigma}{4 \mathcal{N} }\begin{pmatrix}
   \lambda^2 & 0 \\
   0 & \lambda^2 
  \end{pmatrix}_{sp} \, ,\\
  \sum_{j } \tilde{\Gamma}_{s p j}^{1,1,0} &\simeq \sum_{j \sim h} \Delta_{s p j}^{0,1,1} \simeq \frac{i }{2 \mathcal{N} v_\sigma}\begin{pmatrix}
   m_R^4 & 0 \\
   0 & m_R^4
  \end{pmatrix}_{sp} \simeq \frac{i v_\sigma^3}{8 \mathcal{N} }\begin{pmatrix}
   \lambda^4 & 0 \\
   0 & \lambda^4
  \end{pmatrix}_{sp} \, .
   \end{align}
Finally we obtain the leading order contribution for the 1-loop coupling of the majoron to charged leptons,
\begin{equation}
\label{eq:Majinthybrid}
\mathcal{L}_{\ell\ell J} =\frac{iJ}{32\pi^2 v_\sigma} \bar{\ell}\left[M_\ell ~ \textup{Tr}(y_N \, y_N^\dagger) \, \gamma_5 +2 M_\ell \, \left(y_N \, y_N^\dagger-  y_S \, \Theta \, y_S^\dagger \right) P_L- 2 \, \left(y_N \, y_N^\dagger - y_S \, \Theta \, y_S^\dagger \right)M_\ell P_R\right]\ell \, ,
\end{equation}
where the matrix $\Theta$ is given by
\begin{equation}
    \Theta_{sp} \equiv \frac{(m_R^2)_s}{\left( (m_R^2)_s - m_{\eta^+}^2 \right)^2}  \left( (m_R^2)_s- m_{\eta^+}^2 + m_{\eta^+}^2 \log \frac{m_{\eta^+}^2}{(m_R^2)_s} \right) \delta_{sp} \, .
\end{equation}
The strength of the majoron interactions showcased in Eq.~\eqref{eq:Majinthybrid} is not neutrino mass-suppressed. Indeed, we can consider the limit $\mu_S, v_\chi \to 0$, which leads to $M_\nu \to 0$ in Eq.~\eqref{eq:mnuhybrid}. Even in this unrealistic scenario with massless neutrinos, Eq.~\eqref{eq:Majinthybrid} would lead to a lepton flavor violating majoron with potentially observable signatures.

\subsection{Inverse seesaw with enhanced majoron LFV}
\label{subsec:Inverseenhanced}

\begin{table}[tb!]
\begin{center}
\renewcommand{\arraystretch}{1.35} 
\setlength{\tabcolsep}{5pt} 
\begin{tabular}{| c | c | c |}
\hline
Fields & $SU(2)_L \otimes U(1)_Y$ & $U(1)_L $ \\ 
\hline
$H$ & $ (\textbf{2}, \frac{1}{2}) $ & 0 \\ 
$\sigma$ & $(\textbf{1}, 0)$ & $-1$ \\
\hline
$L$ & $(\textbf{2}, -\frac{1}{2})$ & 1 \\
$N$ & $ (\textbf{1}, 0) $ & 1  \\
$S$ & $ (\textbf{1}, 0) $ & $0$  \\ \hline
\end{tabular}
\end{center}
\caption{Lepton number and gauge electroweak charges of the particles in model $C2b\slashed{\chi}$. This model leads to an inverse seesaw mechanism for neutrino masses.}
\label{tab:C2bnoChi}
\end{table}

The $C2b\slashed{\chi}$ model can be obtained by removing the doublet $\chi$ from the field inventory of model $C2b$. Therefore, the charges of the fields in this model are those shown in Table~\ref{tab:C2bnoChi}. The Yukawa Lagrangian is given by
\begin{equation}
-\L = Y \bar{L} H e_R + y_N \, \bar{L} \tilde{H} N  + \lambda  \, \sigma  \bar{N}^c S +\frac{1}{2} \, \mu_S\, \bar{S}^c S   + \hc \, ,
\end{equation}
which corresponds to that of the previous model after setting $y_S = 0$. One obtains the neutral fermion mass matrix
\begin{equation}
\mathcal{M} = \begin{pmatrix}
    0 & m_D & 0\\
   m_D^T& 0 & m_R \\
    0 & m_R^T & \mu_S
    \end{pmatrix} \, ,
\end{equation}
with $m_D = y_N \, v_H/\sqrt{2}$ and $m_R = \lambda \, v_\sigma/\sqrt{2}$. By assuming the VEV hierarchy $\mu_S \ll v_H\ll v_\sigma$, the neutrino mass matrix is found to be
\begin{equation}
\label{eq:mnuinverse}
M_\nu \approx \frac{v_H^2}{v_\sigma^2} \, y_N \, \left( \lambda^T \right)^{-1} \mu_S \lambda^{-1} y_N^T \, ,
\end{equation}
which is nothing but the well-known neutrino mass formula obtained in the inverse seesaw. The spectrum of the scalar sector of this model can be easily derived by adapting the results from the previous Section. Similarly, the majoron coupling to charged leptons in this model can be obtained simply by
taking the limit $m_{\eta^+} \to \infty$ (or, equivalently, $y_S \to 0$) in Eq.~\eqref{eq:Majinthybrid}. One finds
\begin{equation}
\label{eq:Majininverse}
\mathcal{L}_{\ell\ell J} =\frac{iJ}{32\pi^2 v_\sigma} \bar{\ell}\left[M_\ell ~ \textup{Tr}(y_N \, y_N^\dagger) \, \gamma_5 +2 M_\ell \, y_N \, y_N^\dagger \, P_L- 2 \, y_N \, y_N^\dagger \, M_\ell P_R\right]\ell \, .
\end{equation}
We note once again that the couplings in Eq.~\eqref{eq:Majininverse} do not vanish in the limit $\mu_S \to 0$, which implies sizable majoron LFV rates even in the absence of neutrino masses.

\subsection{Linear seesaw with enhanced majoron LFV}
\label{subsec:Linearenhanced}

\begin{table}[tb!]
\begin{center}
\renewcommand{\arraystretch}{1.35} 
\setlength{\tabcolsep}{5pt} 
\begin{tabular}{| c | c | c |}
\hline
Fields & $SU(2)_L \otimes U(1)_Y$ & $U(1)_L $ \\ 
\hline
$H$ & $ (\textbf{2}, \frac{1}{2}) $ & 0 \\ 
$\chi$ & $ (\textbf{2}, \frac{1}{2}) $ & $-4$ \\ 
$\sigma$ & $(\textbf{1}, 0)$ & $-2$ \\
\hline
$L$ & $(\textbf{2}, -\frac{1}{2})$ & 1 \\
$N$ & $ (\textbf{1}, 0) $ & 1  \\
$S$ & $ (\textbf{1}, 0) $ & $-3$  \\ \hline
\end{tabular}
\end{center}
\caption{Lepton number and gauge electroweak charges of the particles in model $C3$. This model leads to a linear seesaw mechanism for neutrino masses.}
\label{tab:C3}
\end{table}

Let us now consider model $C3$ which, as described in Sec.~\ref{sec:spontaneous}, also features enhanced rates of majoron LFV processes. The charges of the fields under the electroweak and global $U(1)_L$ symmetries are given explicitly in Table~\ref{tab:C3}. The relevant Yukawa Lagrangian is
\begin{equation}
-\L = Y \bar{L} H e_R + y_N \, \bar{L} \tilde{H} N  +y_s \,  \bar{L} \tilde{\chi} S + \lambda  \, \sigma^*  \bar{N}^c S +\frac{1}{2} \, \lambda_N \, \sigma \, \bar{N}^c N   + \hc \, ,
\end{equation}
which leads to the neutral fermion mass matrix
\begin{equation}
\mathcal{M} = \begin{pmatrix}
    0 & m_D & m_L\\
   m_D^T& \mu_N & m_R \\
    m_L^T & m_R^T & 0
    \end{pmatrix} \, ,
\end{equation}
with $m_D = y_N \, v_H/\sqrt{2}$, $m_L = y_S \, v_\chi/\sqrt{2}$, $\mu_N = \lambda_N \, v_\sigma/\sqrt{2}$ and $m_R = \lambda \, v_\sigma/\sqrt{2}$. This texture, assuming $v_\chi \ll v_H \ll v_\sigma$, leads to a linear seesaw mechanism for neutrino masses, with
\begin{equation}
\label{eq:mnulinear}
    M_\nu \approx - m_L \, m_R^{-1} \, m_D^T + \text{tr.} = -\frac{v_\chi v_H}{v_\sigma} \, y_S \, \lambda^{-1} \, y_N^T + \text{tr.}
\end{equation}
One of the matrices $\lambda$ and $\lambda_N$ can be taken to be diagonal without loss of generality by performing adequate rotations of the fields $N$ and $S$. In the scalar sector, the $U(1)_L$ charges of $\chi$ and $\sigma$ are $-4$ and $-2$, respectively. Then, the scalar potential, apart from the usual self-conjugate terms, also includes the terms
\begin{equation}
    V \supset \lambda_{H\chi 2}\, \left(H \chi^\dagger\right) \left(\chi H^\dagger\right) + \lambda_G \, \chi^\dagger H \sigma \sigma + \hc \, .
\end{equation}
Using the same notation for the scalar and pseudoscalar states and assuming the same VEV hierarchy ($v_\chi \ll v_H \ll v_\sigma$) as in Sec.~\ref{subsec:Hybridenhanced}, the mass spectrum of the model is given by:
\begin{itemize}
    \item CP-even scalars: $m_h^2\sim v_H^2$, $m_{{\cal S}_2}^2 \sim v_\sigma^2 v_H/v_\chi$ and $m_{{\cal S}_3}^2 \sim v_\sigma^2$
    \item CP-odd scalars: $m_{G_Z} = m_J = 0$ and $m_{{\cal P}_3}^2= -\lambda_G \frac{v_\sigma^2 v^2}{2 v_H v_\chi} \mathcal{N}^2\sim v_\sigma^2 \, v/v_\chi$ 
    \item Charged scalars: $m_{G_W}=0$ and $m^2_{\eta^+} = -\frac{\lambda_{H\chi 2}}{2}v^2 - \frac{\lambda_G}{2}\frac{v_\sigma^2 v^2}{v_H v_\chi} \sim v_\sigma^2 \, v/v_\chi$ 
\end{itemize}
Here we have introduced the normalization factor
\begin{equation} \label{eq:Nlinear}
\mathcal{N}^2 = 1 + 4 \, \frac{v_H^2 \, v_\chi^2}{v_\sigma^2 \, v^2} \, . 
\end{equation}

Again, as expected, we find a SM-like Higgs, $h$, as well as the usual Goldstone bosons, including the majoron due to the spontaneous breaking of global $U(1)_L$.
The masses of the massive pseudoscalar $\mathcal{P}_3$ and the charged scalar $\eta^+$ lie above the $v_\sigma$ scale by a factor of order $\sqrt{v / v_\chi} \gg 1$ and thus decouple from the theory at lower energies. Therefore, one can easily verify that the diagrams (c) and (d) of Fig.~\ref{fig:diagrams} give subdominant contributions and we have to compute only the gauge boson contributions, diagrams (a) and (b) of the same figure. We start by writing the profile of the majoron $J$ in the basis  $( H^a, \chi^a, \sigma^a)$,
\begin{align}
\label{eq:majoronC3}
    J =  \frac{1}{\mathcal{N}} \left( - 2 \, \frac{v_\chi^2 \, v_H}{v_\sigma \, v^2}, \, 2 \, \frac{v_H^2 \, v_\chi}{v_\sigma \, v^2}, \, 1 \right) \, .
\end{align}
Again, the superscript $a$ denotes the imaginary component, and we have used the normalization factor of Eq.~\eqref{eq:Nlinear}.

Given the hierarchies at play, $v^2 \approx v_H^2$ and the majoron is mainly a singlet, with a small doublet component suppressed by a factor of order $m_\nu / v$ or smaller. In fact, this component leads to a very suppressed tree-level diagonal coupling between the majoron and charged leptons:
\begin{align}
  \mathcal{L}_{\ell\ell J}^{\textup{Tree-Level}} =2 i \frac{v_\chi^2 v_H}{\mathcal{N}v^3 v_\sigma} J \bar{\ell} M_\ell \gamma_5 \ell \sim \frac{m_\nu \, m_\ell \, v_\chi}{v^3} \, J \bar{\ell} \, \gamma_5 \, \ell \, .
\end{align}
However, larger couplings are found at the 1-loop level. To compute these coupling we must determine $\bar{A}$ (see Eq.~\eqref{eq:Abar}) in this model. One finds
\begin{align}
\bar{A} = \frac{1}{2} \begin{pmatrix}
0 & \bar{A}_L \\
\bar{A}_L^T & \bar{A}_H
\end{pmatrix} \, , \, \, \, \, \bar{A}_L = 2i\frac{v_\chi v_H}{\sqrt{2} \mathcal{N} v^2 v_\sigma}\begin{pmatrix}
-v_\chi y_N & v_H y_S 
\end{pmatrix} \, , \, \, \, \, \bar{A}_H = \frac{i}{\sqrt{2} \mathcal{N}} \begin{pmatrix}
    \lambda_N & -\lambda \\
    -\lambda^T & 0
\end{pmatrix} \, ,
\end{align}
and then
\begin{align}
  \sum_{j \sim l } \Gamma_{\beta \alpha j}^{1,0,0} &= \sum_{j \sim l} \tilde{\Gamma}_{\beta \alpha j}^{1,0,0}\simeq  \frac{i}{2 v_\sigma}  \left (m_D m_D^\dagger \right)_{\beta \alpha} \, =  i\frac{v^2}{4 v_\sigma}  \left( y_N \, y_N^\dagger \right)_{\beta \alpha} \, , \\
   \sum_{j \sim h} \Delta_{\beta \alpha j}^{0,1,-1} &= \sum_{j \sim h} \tilde{\Delta}_{\beta \alpha j}^{0,1,-1} \simeq -\frac{i}{ 2v_\sigma}  \left (m_D m_D^\dagger -2 m_D \mu_N^\dagger \mu_N \left( m_R^T\right)^{-1} \left( m_R^*\right)^{-1} m_D^\dagger \right)_{\beta \alpha} \\
   &= -i\frac{v^2}{ 4 v_\sigma}  \left (y_N \, y_N^\dagger -\frac{1}{2} y_N \lambda_N^\dagger \lambda_N \left( \lambda^T\right)^{-1} \left( \lambda^*\right)^{-1} y_N^\dagger \right)_{\beta \alpha} \, .
   \end{align}
Finally, with these results at hand, one can write the leading contribution for the coupling of the majoron to a pair of charged leptons as
\begin{align}
\label{eq:Majintlinear}
\mathcal{L}_{\ell\ell J} = -\frac{iJ}{32\pi^2 v_\sigma} \, \bar{\ell} \, \Bigg\{& M_\ell ~\RE \, \left[\textup{Tr}\left( y_N \,\left(\id_3 -\frac{1}{3}R \right) y_N^\dagger \right) \right] \gamma_5 + M_\ell \, y_N \,\left( 2 \, \id_3 -\frac{5}{12}R \right) y_N^\dagger P_L \nonumber \\
&-  2 y_N \,\left( 2 \, \id_3 -\frac{5}{12}R^\dagger \right) y_N^\dagger \, M_\ell P_R\Bigg\} \, \ell \, ,
\end{align}
where $R \equiv \lambda_N^\dagger \lambda_N \left( \lambda^T\right)^{-1} \left( \lambda^*\right)^{-1}$. As in the previous two example models, the strength of the majoron interactions in Eq.~\eqref{eq:Majintlinear} is not neutrino mass suppressed. Again, considering the limit $v_\chi \to 0$ leads to vanishing neutrino masses ($M_\nu \to 0$ in Eq.~\eqref{eq:mnulinear}), but leaves Eq.~\eqref{eq:Majintlinear} intact.

\subsection{Analysis}
\label{sec:analysis}

We have just examined three models featuring enhanced majoron couplings to charged leptons. To determine whether this enhancement will be phenomenologically significant we will now analyze the current constraints on this coupling and their future improvement prospects. We will also compare them with other typical signals of low-scale seesaws, such as $\mu \to e \gamma$.

\subsubsection{Current and future constraints}
We start by writing down a general majoron interaction Lagrangian with a pair of charged leptons,
\begin{equation} \label{eq:majoronpheno}
\mathcal{L}_{\ell\ell J} = J \bar{\ell}\left( S_L P_L + S_R P_R\right) \ell + \hc  =J \bar{\ell}\left( S P_L + S^\dagger P_R\right) \ell \, ,
\end{equation}
with $S = S_L + S_R^\dagger$. There are stringent constraints on both diagonal and off-diagonal majoron couplings to
charged leptons~\cite{Escribano:2020wua}. The flavor conserving couplings are constrained by energy loss mechanisms in astrophysical observations~\cite{DiLuzio:2020wdo,Calibbi:2020jvd,Croon:2020lrf, Caputo:2021rux} and yield
\begin{align}
\label{eq:stellarcooling}
    \left| \text{Im} \left(S_{ee}^\text{exp}\right) \right| &< 2.1 \times 10^{-13} \, , \\
    \left| \text{Im} \left(S_{\mu \mu}^\text{exp}\right) \right| &< 3.1 \times 10^{-9} \, .
    \label{eq:stellarcooling2}
\end{align}
The presence of non-zero off-diagonal couplings in Eq.~\eqref{eq:majoronpheno} allows the non-standard decay $\mu^+ \to e^+ \, J$. In particular, we find 
\begin{equation} \label{eq:gamma}
\Gamma(\ell_\alpha \to \ell_\beta \, J) = \frac{m_{\ell_\alpha}}{32 \, \pi} \, \left| \widetilde S^{\alpha \beta} \right|^2 \, ,
\end{equation}
where we have defined 
\begin{equation}
  \left| \widetilde S^{\alpha \beta} \right| = \left( \left| S^{\alpha \beta }_L \right|^2 + \left| S^{\alpha \beta }_R \right|^2 \right)^{1/2} \, .
\end{equation}

The best limits on this process were obtained at TRIUMF~\cite{Jodidio:1986mz}. Taking into account all
possible chiral structures for the majoron coupling, one can estimate
the limit~\cite{Hirsch:2009ee}
\begin{equation} \label{eq:majoronlimit}
    \BR \left( \mu \to e \, J \right) \lesssim 10^{-5} \, ,
\end{equation}
which in turn implies
\begin{equation} \label{eq:mulim}
  \left| \widetilde S^{\mu e} \right| < 5.3 \times 10^{-11} \, .
\end{equation}
Finally, the currently best experimental limits on $\tau$ decays
including majorons were set by the Belle II
collaboration~\cite{Belle-II:2022heu}. They can be used to derive the
bounds
\begin{equation} \label{eq:taulim}
  \begin{split}
    & \left| \widetilde S^{\tau e} \right| < 3.5 \times 10^{-7} \, , \\
    & \left| \widetilde S^{\tau \mu} \right| < 2.7 \times 10^{-7} \, .
  \end{split}
\end{equation}
Future experiments such as Mu3e~\cite{Hesketh:2022wgw,Perrevoort:2024qtc} or COMET \cite{COMET:2018auw,Xing:2022rob} will improve these constraints. In particular, the expected future sensitivities for a massless majoron are expected to be
\begin{align}
    \BR\left(\mu \rightarrow e \, J \right)_\text{Mu3e} &\lesssim 6 \,\cdot \, 10^{-7} \,  \\
    \BR\left(\mu \rightarrow e \, J \right)_\text{COMET} &\lesssim 4.6 \,\cdot \, 10^{-9} \, 
\end{align}
Other muon decays with a majoron in the final state that can be used to probe a flavor violating coupling are $\mu \to e J \gamma$~\cite{Hirsch:2009ee,Jho:2022snj,Herrero-Brocal:2023czw}, $\mu \to eee J$~\cite{Knapen:2023zgi} and $\mu \to e J J$.

A typical signal of many BSM models, and in particular low-scale seesaws, is the lepton flavor violating process $\mu \to e \gamma$. The MEG collaboration reported~\cite{MEG:2016leq}
\begin{equation}
    \BR\left(\mu \rightarrow e \, \gamma \right)_\text{MEG} \lesssim 4.2 \, \cdot \, 10^{-13} \, .
\end{equation}
This bound will be improved by MEG-II~\cite{MEGII:2021fah} to
\begin{equation}
        \BR\left(\mu \rightarrow e \, \gamma \right)_\text{MEG-II} \lesssim 6 \, \cdot \, 10^{-14} \, .
\end{equation}

\subsubsection{Phenomenology}

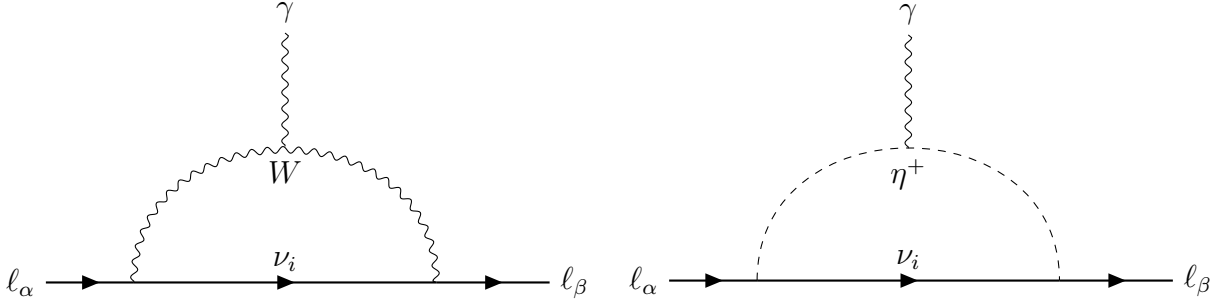
\begin{figure}[tb!]
\centering
\begin{minipage}{.5\textwidth}
  \centering
  \begin{tikzpicture}
    \begin{feynman}
      \vertex (a) {\(\ell_\alpha\)};
      \vertex [right=1.5cm of a] (b); 
      \vertex [right=4cm of b] (c);
      \vertex [right=1.5cm of c] (d) {\(\ell_\beta\)};
      \vertex [above right=1.8cm and 2cm of b] (top);
      \vertex [above=1.5cm of top] (gamma) {\(\gamma\)};

      \diagram* {
        (a) -- [fermion, thick] (b) -- [fermion, thick, edge label=\(\nu_i\)] (c) -- [fermion, thick] (d),
        (b) -- [boson, half left, looseness=1.5, edge label'=\(W\), midway] (c),
        (top) -- [photon] (gamma),
      };
    \end{feynman}
  \end{tikzpicture}
\end{minipage}%
\begin{minipage}{.5\textwidth}
  \centering
  \begin{tikzpicture}
    \begin{feynman}
      \vertex (a) {\(\ell_\alpha\)};
      \vertex [right=1.5cm of a] (b); 
      \vertex [right=4cm of b] (c);
      \vertex [right=1.5cm of c] (d) {\(\ell_\beta\)};
      \vertex [above right=1.75cm and 2cm of b] (top);
      \vertex [above=1.5cm of top] (gamma) {\(\gamma\)};

      \diagram* {
        (a) -- [fermion, thick] (b) -- [fermion, thick, edge label=\(\nu_i\)] (c) -- [fermion, thick] (d),
        (b) -- [scalar, half left, looseness=1.5, edge label'=\(\eta^+\), midway] (c),
        (top) -- [photon] (gamma),
      };
    \end{feynman}
  \end{tikzpicture}
\end{minipage}
\caption{Feynman diagrams relevant for the muon decay $\mu \rightarrow e \gamma$. The fermion mediators include the light and heavy neutral fermions. \textbf{Left diagram:} mediation by \(W\) boson. \textbf{Right diagram:} mediation by scalar \(\eta^+\).}
\label{fig:muegamma}
\end{figure}

We now proceed to analyze the majoron phenomenology of our selected models $C2b$, $C2b\slashed{\chi}$ and $C3$. In all of them we can estimate
\begin{align} \label{eq:gammaEst}
\Gamma(\ell_\alpha \to \ell_\beta \, J) \sim \frac{m_{\ell_\alpha}^3}{v_\sigma^2} \, .
\end{align}
On the other hand, the $\ell_\alpha \rightarrow \ell_\beta \gamma$ process will be mediated in our general setup by the $W$ boson and the charged scalar $\eta^+$ whenever it is present, see Fig.~\ref{fig:muegamma}. While the exact expressions are well-known, see for instance~\cite{Lavoura:2003xp}, we can again estimate

\begin{equation}
    \Gamma(\ell_\alpha \to \ell_\beta \, \gamma) \sim  \frac{m_{\ell_\alpha}^5}{m_R^4}
\end{equation}
where $m_R$ is the mass of the heavy neutral fermion running in the loop. In all the models considered, this mass will be approximately given by the lepton number breaking scale $v_\sigma$ times some Yukawa coupling $\lambda$. Then, the ratio between the branching ratio of both processes will scale as
\begin{equation}
\label{eq:ratio}
    \frac{\BR\left(\ell_\alpha \rightarrow \ell_\beta \gamma \right)}{\BR\left(\ell_\alpha \rightarrow \ell_\beta J \right)} \sim  \left(\frac{{m_{\ell_\alpha}}}{m_R}\right)^2 \lambda^{-2} \, ,
\end{equation}
and hence one expects $\BR\left(\ell_\alpha \rightarrow \ell_\beta \gamma \right) \ll \BR\left(\ell_\alpha \rightarrow \ell_\beta J \right)$, except possibly when $\lambda \ll 1$. For instance, for $\ell_\alpha = \mu$ and $m_R \sim 1$ TeV,  $\BR\left(\ell_\alpha \rightarrow \ell_\beta \gamma \right) \sim \BR\left(\ell_\alpha \rightarrow \ell_\beta J \right)$ would require $\lambda \lesssim 10^{-4}$. In summary, the key signature in these models is the flavor violating majoron emission in lepton decays. 

The previous analysis is overly simplistic, as it only considers the relevant energy scales and their hierarchy. Therefore, a rigorous numerical exploration that takes into account the various numerical factors arising from the loops and the freedom in the scale of the Yukawa couplings is required to validate it. As we proceed to show, a correct treatment proves that the parameter space where the process $\mu \rightarrow e \gamma$ would be observed before $\mu \rightarrow e J$ is extremely limited, particularly needing small Yukawa couplings, thus moving away from the motivation for low-scale seesaws. In other words, while these models could feature sizeable $\mu \rightarrow e \gamma$ decays, as expected in similar low-scale seesaw models, most of the parameter space in which this happens would lead to majoron interactions which are already ruled out by astrophysical observations and cLFV experiments.

Since our analysis holds true irrespective of the charged scalar $\eta^+$ mass, we can study the majoron-charged lepton phenomenology of the three models in the limits $\kappa \gg v_\sigma v_\chi/v$, leading to $\Theta \rightarrow 0$ in Eq.~\eqref{eq:Majinthybrid}, and $\lambda_N \rightarrow 0$ in Eq.~\eqref{eq:Majintlinear}. In both simplifying scenarios the neutrino mass remains unaffected and the majoron-charged lepton interaction becomes equal for the three models, $C2b$, $C2b\slashed{\chi}$ and $C3$,~\footnote{Note that there is a non-physical global sign between the results in  $C2b$ and $C2b\slashed{\chi}$ and the one in $C3$ that appears due to the lepton charge choice.} i.e.
\begin{equation}
\label{eq:Majinttoy}
\mathcal{L}_{\ell\ell J} =\frac{iJ}{32\pi^2 v_\sigma} \bar{\ell}\left[M_\ell ~ \textup{Tr}(y_N \, y_N^\dagger)\gamma_5 +2 M_\ell \, y_N \, y_N^\dagger  \, P_L- 2 \,y_N \, y_N^\dagger M_\ell \, P_R\right]\ell \, .
\end{equation}
Comparing this expression with the generic Lagrangian in Eq.~\eqref{eq:majoronpheno}, we identify
\begin{align}
    S_L &=  \frac{i}{64 \pi^2 v_\sigma} \left[2 M_\ell \, y_N y_N^\dagger - M_\ell \, \text{Tr}\left(y_N y_N^\dagger\right)\right]\, , \label{eq:SL1} \\
    S_R & = S_L^\dagger \, . \label{eq:SL2}   
\end{align}
For any given Yukawa matrix $y_N$ it is possible to find suitable $\mu_S$, $y_S$ and $v_\chi$ that fit neutrino data~\cite{Casas:2001sr,Cordero-Carrion:2018xre,Cordero-Carrion:2019qtu}. Out of the astrophysical constraints, we expect the constraint of Eq.~\eqref{eq:stellarcooling} to be more stringent than the one of Eq.~\eqref{eq:stellarcooling2}. Indeed, in order for this not to be the case we would need
\begin{equation}
    \frac{S_{\mu \mu}}{S_{ee}} \gtrsim 10^4 \, ,
\end{equation}
but from Eqs.~\eqref{eq:SL1}-\eqref{eq:SL2}, and taking into account that $S = S_L + S_R^\dagger$, one finds
\begin{equation}
    \frac{S_{\mu \mu}}{S_{ee}} = \frac{m_\mu}{m_e} \times \frac{-(y_N \, y_N^\dagger)_{11}+(y_N \, y_N^\dagger)_{22} - (y_N \, y_N^\dagger)_{33}}{(y_N \, y_N^\dagger)_{11}-(y_N \, y_N^\dagger)_{22} - (y_N \, y_N^\dagger)_{33}} \, ,
\end{equation}
which implies that fine-tuned Yukawas couplings are required to cancel the contribution to $S_{ee}$. If instead we assume the Yukawa couplings $y_N$ and $\lambda$ to be of $\mathcal{O}(1)$, we can use Eqs.~\eqref{eq:gamma} and \eqref{eq:majoronlimit}, as well as the couplings in Eqs.~\eqref{eq:SL1}-\eqref{eq:SL2}, to derive a lower limit for the lepton number breaking scale $v_\sigma$,
\begin{align}
    v_\sigma > \frac{3 \, m_e}{32 \pi^2} \frac{1}{\left| \text{Im} \left(S_{ee}^\text{exp}\right) \right|} &\sim 10^4 \, \text{TeV} \, ,
\end{align}
and an estimate for the branching ratios of the flavor violating processes $\mu \to e J$ and $\mu \to e \gamma$ obtained for that scale,
\begin{align}
    \BR\left(\mu \rightarrow e J\right) &\approx 10^ {-5} \, , \\
    \BR\left(\mu \rightarrow e \gamma\right) &\approx 10^{-23} \, .
\end{align}

\begin{figure}[t!]
\centering
\includegraphics[height=6.2cm]{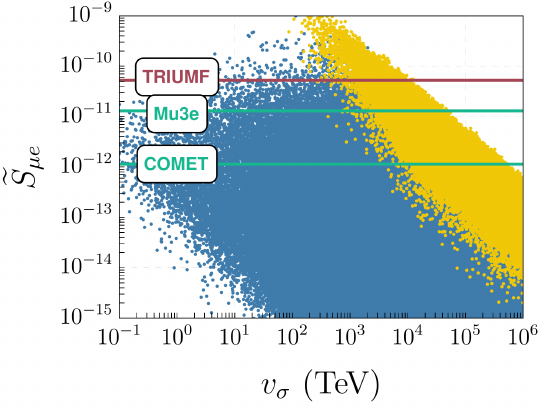}
\includegraphics[height=6.2cm]{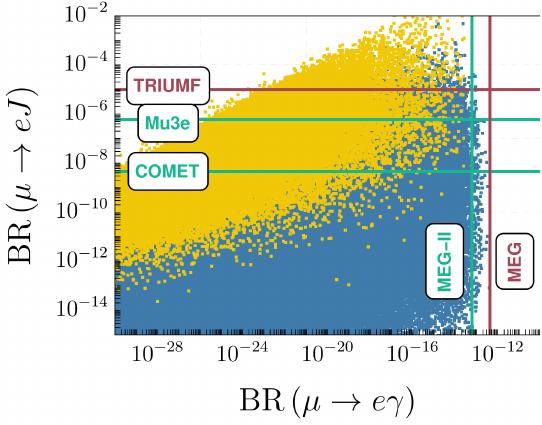}
\caption{Lepton flavor violation predictions in the selected models. \textbf{Left panel:} Relationship between the lepton number breaking scale $v_\sigma$ and the flavor violating coupling $\widetilde S_{\mu e}$. \textbf{Right panel:} Comparison between the branching ratios of the non-standard muon decays with a majoron or a photon in the final state. In both panels we are imposing the seesaw condition $(M_D \cdot M_F^{-1})_{ij} < 10^{-2}$, the astrophysical constraints of Eqs.~\eqref{eq:stellarcooling} and \eqref{eq:stellarcooling2}, as well as correct neutrino masses and mixing. The orange points have Yukawa couplings $y_N, \lambda \sim \mathcal{O}(1)$, while the blue points allow for more freedom, with the Yukawa couplings taking values in a wider range, $y_N, \lambda \sim \mathcal{O}(10^{-3}-1)$.}
\label{fig:Semu}
\end{figure}

This rather strong conclusion can be relaxed by allowing smaller Yukawa couplings. Indeed, in the left panel of Fig.~\ref{fig:Semu} we show the relation between $\widetilde{S}_{\mu e}$ and the lepton number breaking scale $v_\sigma$, while the right panel of this figure shows a comparison between the branching ratios of the two lepton flavor violating processes discussed above. In the numerical scan leading to this figure we have imposed the seesaw condition $(M_D \cdot M_F^{-1})_{ij} < 10^{-2}$, the astrophysical constraints of Eqs.~\eqref{eq:stellarcooling} and \eqref{eq:stellarcooling2}, as well as neutrino masses and mixing in agreement with current data~\cite{deSalas:2020pgw}. In addition, in both panels we have explicitly shown the predictions of the model when the Yukawa couplings are taken to be of $\mathcal{O}(1)$ (orange points) and when this assumption is relaxed and smaller Yukawa couplings are allowed (blue points). First of all,  we observe on the left side of the figure that the current limit on BR($\mu \to e J$) from TRIUMF (see Eq.~\eqref{eq:majoronlimit}) already restricts the parameter space of the model, ruling out some points not excluded by the other constraints (Eq.~\eqref{eq:stellarcooling} being the most important one). We also find that the decay $\mu \to e J$ can be used to test the model at Mu3e even if the lepton number breaking scale $v_\sigma$ is as high as $\sim 4 \cdot 10^4$ TeV. As already explained, in this model this decay has much better chances of being observed than the more conventional $\mu \rightarrow e \gamma$. This is illustrated on the right side of Fig.~\ref{fig:Semu}, where we see that only a very small fraction of the points resulting from our numerical scan will be tested by MEG-II, even though this experiment is sensitive to much smaller branching ratios than those tested by Mu3e for $\mu \to e J$. Importantly, and consistently with our previous considerations, the very few points that will be tested by MEG-II involve small Yukawa couplings, hence departing from the main motivation for low-scale seesaw models.

\begin{figure}[t!]
\centering
\includegraphics[height=6.1cm]{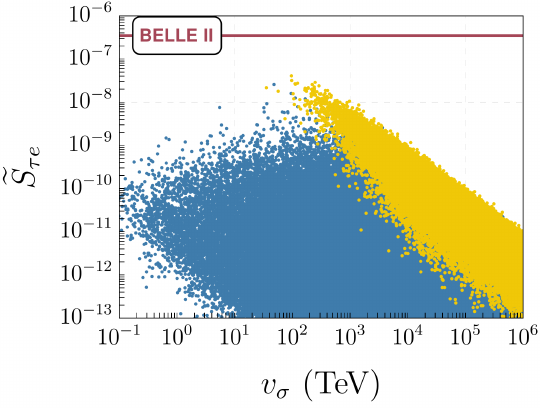}
\includegraphics[height=6.1cm]{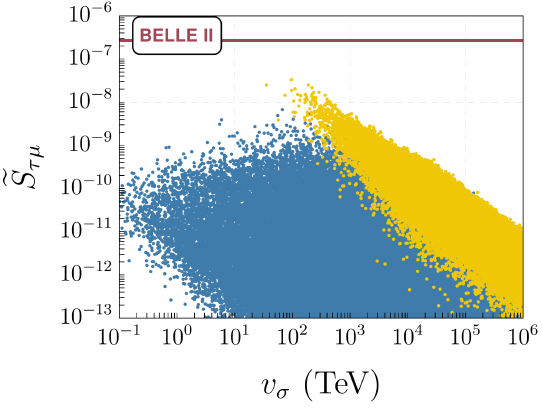}
\caption{Lepton flavor violation predictions for $\tau$ decays in the selected models. In both panels we are imposing the seesaw condition $(M_D \cdot M_F^{-1})_{ij} < 10^{-2}$, the astrophysical constraints of Eqs.~\eqref{eq:stellarcooling} and \eqref{eq:stellarcooling2}, as well as correct neutrino masses and mixing. The orange points have Yukawa couplings $y_N, \lambda \sim \mathcal{O}(1)$, while the blue points allow for more freedom, with the Yukawa couplings taking values in a wider range, $y_N, \lambda \sim \mathcal{O}(10^{-3}-1)$. The rates for both processes $\tau \to J \, e/\mu $ are predicted to be several orders of magnitude below observational limits.}
\label{fig:taudecays}
\end{figure}

Since our setup is flavor-blind, the elements of $\widetilde{S}_{\alpha \beta}/m_{\ell_\alpha}$ are expected to be of similar order (except in the presence of fine-tuned Yukawa couplings). However, $\tau$ decays are experimentally less constrained than $\mu$ decays, see Eqs.~\eqref{eq:majoronlimit}-\eqref{eq:taulim}, and as such are not expected to be observed in the near future. This feature is showcased in Fig.~\ref{fig:taudecays}. Combining our framework with flavor symmetries may further enhance the $\tau$ decays into observable rates, but we do not explore this option here.

We have compared two of the most promising processes for the search for new physics in the flavor sector, namely $\mu \to e \gamma$ and $\mu \to e J$. Nevertheless, we would like to note that these processes are not the only possible ones. Specifically, processes such as $\mu \to e \, JJ$ are possible. However, this 3-body decay is not easy to observe experimentally due to its kinematic conditions and therefore we do not explore it further.


\section{Summary and discussion}
\label{sec:summary}
Models belonging to the Type-I Seesaw family are among the most promising for explaining neutrino masses and their scale. In these models, lepton number is broken explicitly or spontaneously, resulting in a Majorana mass matrix for neutrinos. At first, we focused on the mass matrix, overlooking the origin of the $U(1)_L$ breaking. Using the general seesaw expansion, we derived a comprehensive formula for neutrino masses in the Type-I Seesaw family. This formula reproduces the known results of models such as the type-I, linear, or inverse seesaws, and enables us to identify hybrid models or those with less-known hierarchies.

Regarding the $U(1)_L$ breaking origin, we observed that in the first scenario, with explicit breaking, the differences between models of the family are spurious since lepton number is not a good symmetry, allowing us to describe all models with the same Lagrangian ---the one of the conventional type-I seesaw--- with just different textures. However, when $U(1)_L$ is spontaneously broken, models become distinguishable, and we can no longer describe them with only one Lagrangian. With this realization, we analytically derived all the different minimal models of the Type-I Seesaw family with SSB of $U(1)_L$. In this case, a Goldstone boson arises in the spectrum, the majoron, providing a clear signal of these models and allowing us to distinguish between them based on their phenomenology. We systematically classified this in Section~\ref{sec:spontaneous}, demonstrating that while in most models the majoron couplings to charged leptons are suppressed by neutrino masses, one also finds some models where this interaction is enhanced. Finally, we considered some example models of the latter type and analyzed their phenomenology, finding a promising signal: although we may not observe the usual flavor violating process $\mu \rightarrow e \gamma$, the exotic $\mu \rightarrow e J$ decay may be within the reach of near-future experiments, even for relatively high lepton number breaking scales.

Our work explores both established models as well as novel ones. Moreover, it provides a comprehensive framework for working with models belonging to the Type-I Seesaw family. While our focus has been on the minimal realization of the family, the formulas for majoron-charged leptons interactions provided in Appendix~\ref{app:coupling} open up possibilities for exploring promising non-minimal realizations. Additionally, these formulas are useful not only for scenarios with exact SSB of $U(1)_L$, but also for cases with soft symmetry breaking, where the majoron is not strictly massless but rather light. This allows for the application of our framework to versions featuring a massive majoron, of potential interest in cosmology and low-energy experiments.

\section*{Acknowledgements}

Work supported by the Spanish grants PID2020-113775GB-I00
(AEI/10.13039/501100011033) and CIPROM/2021/054 (Generalitat
Valenciana). The work of AHB is supported by the grant
No. CIACIF/2021/100 (also funded by Generalitat Valenciana).  AV
acknowledges financial support from MINECO through the Ramón y Cajal
contract RYC2018-025795-I.

\appendix

\section{Models in the Type-I Seesaw family}
\label{sec:app1}

Based on the dependence of the neutrino mass matrix on the relevant physical scales, we consider three generic mass generation mechanisms:
\begin{itemize}
    \item {\bf Type-I seesaw}: if $\displaystyle M_\nu \sim \frac{m_1^2}{M}$, with $m \ll M$,
    \item {\bf Inverse seesaw}: if $\displaystyle M_\nu \sim \frac{m_1^2 m_2}{M^2}$, with $m_2 \ll m_1 \ll M$,
    \item {\bf Linear seesaw}: if $\displaystyle M_\nu \sim \frac{m_1 m_2}{M}$ with $m_2 \ll m_1 \ll M$.
\end{itemize}
There are many models in the Type-I Seesaw family, due to the many different possible hierarchies among the blocks in the $M_D$ and $M_F$ matrices. They can be classified according to these hierarchies. We find three possible cases.

\subsubsection*{Case 1: No hierarchy among the blocks in $\boldsymbol{M_F}$}

If the blocks in the $M_F$ matrix are all of the same order, $m_R \sim \mu_N \sim \mu_S \sim \Lambda_H$, where $\Lambda_H$ is a high-enery scale, the neutrino mass matrix can be written as
\begin{equation}
   M_\nu = c_1 \, m_D \, \Lambda_H^{-1} m_D^T + c_2 \, m_D \, \Lambda_H^{-1} m_L^T \, + c_3 \, m_L \, \Lambda_H^{-1} m_L^T + c_4 \, m_L \, \Lambda_H^{-1} m_D^T + \O(\varepsilon^2) \, ,
\end{equation}
where $c_i$, with $i=1, \dots ,4$, are constants. It is clear that one finds a {\bf type-I seesaw}, regardless of the hierarchy between $m_D$ and $m_L$.

\subsubsection*{Case 2: $\boldsymbol{m_R \gg \mu_N \left(m_R^T \right)^{-1} \mu_S}$}

In this case the neutrino mass matrix can be expressed as
\begin{align}
    M_\nu =   &- m_D \, \left( m_R^T \right)^{-1} \mu_S \, m_R^{-1} m_D^T - m_L \, m_R^{-1} \mu_N \left( m_R^T\right)^{-1} \, m_L^T \nonumber \\
    & + m_D \left( m_R^T \right)^{-1}  \, m_L^T \,  + m_L \, m_R^{-1} m_D^T + \O(\varepsilon^2) \, .
\end{align}
Then, depending on the relative scale of $m_D$ and $m_L$ and their hierarchies with the blocks in $M_F$, we can classify the resulting models as follows:
\begin{enumerate}
\item [\bf 2.1] If $\displaystyle \, \frac{m_L}{m_D} \gg \frac{\mu_S}{m_R} \, \, , \, \, \frac{m_L}{m_D} \ll \frac{m_R}{\mu_N} \, \, \text{and} \, \, m_L \ll m_D \, \, \text{or}\, \,  m_D \ll m_L \, $: {\bf linear seesaw}
\item [\bf 2.2] If $\displaystyle \, \frac{m_L}{m_D} \ll \frac{\mu_S}{m_R} \, \, , \, \, \frac{m_L^2}{m_D^2} \ll \frac{\mu_S}{\mu_N} \, \, \text{and}\, \, \mu_S \ll m_R \, $: {\bf inverse seesaw} (Dirac mass term: $m_D$)
\item [\bf 2.3] If $\displaystyle \, \frac{m_L}{m_D} \ll \frac{m_R}{\mu_N} \, \, , \, \, \frac{m_L^2}{m_D^2} \gg \frac{\mu_S}{\mu_N} \, \, \text{and}\, \,  \mu_N \ll m_R \, $: {\bf inverse seesaw} (Dirac mass term: $m_L$)
\item [\bf 2.4] Otherwise: {\bf type-I seesaw}
\end{enumerate}

\subsubsection*{Case 3: $\boldsymbol{m_R \ll \mu_N \left(m_R^T \right)^{-1} \mu_S}$}

In this case the neutrino mass matrix can be written as
\begin{align}
    M_\nu =  m_D \, \mu_N^{-1} m_D^T + m_L \, \mu_S^{-1} m_L^T - m_D \, \mu_N^{-1} \, m_R \, \mu_S^{-1} \, m_L^T - m_L \, \mu_S^{-1} \, m_R^T \, \mu_N^{-1} \, m_D^T\, + \O(\varepsilon^2).
\end{align}
Again, depending on the relative scale of $m_D$ and $m_L$ and their hierarchies with the blocks in $M_F$, the resulting models can be classified as follows:
\begin{enumerate}
\item [\bf 3.1] If $\displaystyle \, \frac{m_L}{m_D} \ll \frac{m_R}{\mu_N} \, \, , \, \, \frac{m_L}{m_D} \gg \frac{\mu_S}{m_R} \, \, \text{and} \, \, m_L \ll m_D \, \, \text{or} \, \, m_D \ll m_L \, $: {\bf linear seesaw}
\item [\bf 3.2] If $\displaystyle \, \frac{m_L}{m_D} \ll \frac{m_R}{\mu_N} \, \, , \, \, \frac{m_L}{m_D} \gg \frac{\mu_S}{m_R} \, \, \text{and} \, \, m_R \ll \mu_S  \, \, \text{or} \, \, m_R \ll \mu_N \, $: {\bf inverse seesaw} (Dirac mass term: $m_D$)
\item [\bf 3.3] Otherwise: {\bf type-I seesaw}
\end{enumerate}

\section{The majoron coupling to charged leptons}
\label{app:coupling}

Reference~\cite{Herrero-Brocal:2023czw} provides general analytical expressions for the 1-loop coupling of the majoron to a pair of charged leptons which can, in general, be written as~\cite{Escribano:2020wua}
\begin{equation} \label{eq:llJ}
\mathcal{L}_{\ell \ell J} = J \, \bar{\ell}_\beta \left( S_L^{\beta \alpha} \, P_L + S_R^{\beta \alpha} \, P_R \right) \ell_{\alpha} + \hc = J \, \bar{\ell}_\beta \left[ S^{\beta \alpha} \, P_L + \left( S^{\alpha \beta} \right)^* \, P_R \right] \ell_{\alpha} \, ,
\end{equation}
where $P_{L,R} = \frac{1}{2} \left( 1 \mp \gamma_5 \right)$ are the usual chiral projectors while $\ell_{\alpha,\beta}$ are the charged leptons, with $\alpha,\beta$ two generation indices. In the Type-I Seesaw family $S$ is given by
\begin{equation} \label{eq:Scoup}
S^{\beta \alpha} = \frac{1}{8 \pi^2}\left( \delta^{\beta \alpha} \, \Gamma_Z^\alpha +  L_W^{\beta \alpha} +  L_{\eta^+}^{\beta \alpha}+  L_S^{\beta \alpha}\right) \, .
\end{equation}
In this equation, each term represents the contribution of one of the Feynman diagrams shown in Fig.~\ref{fig:diagrams}. These contributions, expressed as functions of certain general matrices and loop functions, are provided in~\cite{Herrero-Brocal:2023czw}. Here, we present them for the sake of completeness. They are given by
\begin{align}
\Gamma_Z^\alpha &= \, i  \, \frac{m_{\ell_\alpha}}{v^2} \, \textup{Im}\left[ \sum_{s=1}^3 \left( \sum_{ j \sim l }\frac{\tilde{\Gamma}_{ssj}^{1,0,0}+\Gamma_{ssj}^{1,0,0}}{6} -\sum_{j\sim h}\frac{\tilde{\Delta}_{ssj}^{0,1,-1}+\Delta_{ssj}^{0,1,-1}}{3} \right)  \right]\, , \label{eq:MZ1} \\
%
  L_W^{\beta \alpha} &= \frac{2 \, m_{\ell_\beta}}{v^2} \, \left[  \sum_{j \sim l}\left( \frac{\Gamma_{\alpha \beta j}^{1,0,0 \, *}}{12}+ \frac{2}{3} \Gamma_{\beta \alpha j}^{1,0,0} \right) - \sum_{j\sim h} \left( \frac{\tilde{\Delta}_{\alpha \beta j}^{0,1,-1 \, *}}{6} +\frac{7}{12}\tilde{\Delta}_{\beta \alpha j}^{0,1,-1} \right)\right]  \, , \label{eq:MW} \\
%
%
L_S^{\beta \alpha} &= \frac{\omega}{4 \left( m_\rho^2 - m_\sigma^2 \right)^2} \Bigg\{ \left( G \, C^\dagger \, M_\ell \right)_{\beta \alpha} \left( -m_\rho^2 + m_\sigma^2 +m_\rho^2 \log \frac{m_\rho^2}{m_\sigma^2} \right) \nonumber \\ 
&- \left( M_\ell G^\dagger \, C  \right)_{\beta \alpha} \left( -m_\rho^2 + m_\sigma^2 +m_\sigma^2 \log \frac{m_\rho^2}{m_\sigma^2} \right) 
-2 \left( G \, M_\ell \, C \right)\left( m_\rho^2 - m_\sigma^2 \right) \log \frac{m_\rho^2}{m_\sigma^2}\Bigg\} \, , \\
%
%
  L_\eta^{\beta \alpha} &= \, m_{\ell_\beta} \left[ \bar{D}_R^{\beta p}\left(\bar{D}_R^{\alpha s}\right)^* \left(L_\eta^{RR}\right)^*_{sp} - \left(\bar{D}_R^{\alpha p}\right)^* \bar{D}_R^{\beta s} \left(\widetilde L_\eta^{RR}\right)_{sp} \right] \, ,
\end{align}
where in the last equation we have defined
\begin{align}
  L_\eta^{RR} =& \, f_7 \, \sum_{j\sim h}\Delta^{0,1,-1}_{spj}+\sum_j \left( f_8 \, \Delta^{0,1,1}_{spj}-F_{5,7}  \,\tilde{\Gamma}^{1,0,0}_{spj} -F_{6,8} \, \tilde{\Gamma}^{1,1,0}_{spj} \right) \nonumber \\
  &+ \sum_{j\sim l} \left( F_{5,7,-1} \, \tilde{\Gamma}^{1,0,0}_{spj}+F_{6,8,-2} \, \tilde{\Gamma}^{1,1,0}_{spj}\right)  \, , \label{eq:LRR}  \\
  \widetilde L_\eta^{RR} =& L_\eta^{RR} \, \left( f_{\left(1, \, 2,\, 5, \, 6, \,7, \, 8 \right) } \leftrightarrow f_{\left(3, \, 4,\, 9, \, 10, \,15, \, 16 \right) },\: f_{13} \leftrightarrow F_{1, -3}, \: f_{14} \leftrightarrow F_{2, -4} \right) \label{eq:LRRtilde} \, ,
\end{align}
and where, following the conventions of~\cite{Herrero-Brocal:2023czw}, $\omega$ denotes the coupling between $J {\cal P}_k {\cal S}_k$, $C$ represents the coupling between charged leptons and ${\cal S}_k$, and $G$ is the coupling between the charged leptons and ${\cal P}_k$. The loop functions $f$ and $F$ are also provided in~\cite{Herrero-Brocal:2023czw}. We note that in these expressions there are sums that extend over all states, or only over the light ($j \sim l$) or heavy ($j \sim h$) ones. The precise definitions of the $\Gamma$, $\tilde{\Gamma}$, $\Delta$, and $\tilde{\Delta}$ matrices are also given in this reference. One can particularize them for the Type-I Seesaw family and find the combinations that appear in Eqs.~\eqref{eq:MZ1}-\eqref{eq:LRRtilde}. The sums relevant for the gauge boson contributions are given by
\begin{align}
  \sum_{ j \sim l } \Gamma_{\beta \alpha j}^{1,0,0 } &= \sum_{ j \sim l } \tilde{\Gamma}_{\alpha \beta j}^{1,0,0 }=\frac{1}{2} \bar{A}_L M_D^\dagger -\frac{1}{2} \left( M_D \, M_F^{-1} \, \bar{A}_H \, M_D^\dagger \right)_{\beta \alpha} \, , \\
  \sum_{ j \sim h} \tilde{\Delta}_{ \beta \alpha j}^{0,1,-1} &=\sum_{ j \sim h} \Delta_{ \alpha \beta j}^{0,1,-1} = \frac{1}{2}M_D \, \left(M_D^\dagger \bar{A}_L + M_F^\dagger \bar{A}_L^T \, M_D^* \, ( M_F^\dagger )^{-1} + M_F^\dagger\bar{A}_H\,  \right) M_F^{-1}\, ( M_F^\dagger )^{-1}  \, M_D^\dagger \, ,
\end{align}
whereas the sums relevant for the triangle diagram with the $\eta$ charged scalar are
\begin{align}
  \sum_{ j } \tilde{\Gamma}_{s p j}^{1,0,0 } &=\frac{1}{2} \left( M_D^\dagger \bar{A}_L + M_F^\dagger \, \bar{A}_H  \right)_{s p} \, , \\
  \sum_{ j } \tilde{\Gamma}_{s p j}^{1,1,0 } &= \frac{1}{2}\left( M_F^\dagger \, M_F \, M_F^\dagger \, \bar{A}_H  \right)_{s p} \, , \\
  \sum_{ j } \Delta_{ s p j}^{0,1,1} &= \frac{1}{2}\left(M_D^\dagger \bar{A}_L M_F^\dagger M_F + M_F^\dagger \bar{A}_L^T \, M_D^* M_F  + M_F^\dagger \bar{A}_H \, M_F^\dagger  M_F \right)_{sp}  \, ,\\
  \sum_{ j \sim h} \Delta_{ s p  j}^{0,1,-1} &=\frac{1}{2} \left[ M_F^{-1} \left( (M_F^\dagger)^{-1} M_D^\dagger \bar{A}_L M_F^\dagger  +  \bar{A}_L^T \, M_D^*  + \bar{A}_H \, M_F^\dagger \right) M_F \right]_{sp}  \, .
\end{align}
Here we have used the fact that the charged scalar does not couple to the charged leptons and SM neutrinos. The amplitudes in these expressions depend on the couplings entering the loops, namely the majoron coupling to a pair of neutral fermions, $\bar{A}$, and the charged scalar coupling to a neutral lepton and a charged lepton, $\bar{D}_{L,R}$.
%
%
These are given in the gauge basis and can be readily computed for all variants in the Type-I Seesaw family. Let us consider the general Lagrangian
\begin{equation}
\label{eq:alphasgeneral}
-\L = y_N \, \bar{L} \tilde{H} N  +y_s \,  \bar{L} \tilde{\chi} S + \lambda  \, \sigma_{NS} \, \bar{N}^c S + \frac{1}{2} \lambda_N \, \sigma_N\, \bar{N}^c N  + \frac{1}{2} \lambda_S \, \sigma_S \bar{S}^c S + \hc \, ,
\end{equation}
which reduces to the models discussed in Sec.~\ref{sec:spontaneous} by properly matching the scalar fields to those in each model. Since we introduce only one singlet scalar $\sigma$ in our field inventory, and this not always participates in all fermion singlet Yukawa terms, $\sigma_{NS}$, $\sigma_N$ or $\sigma_N$ may be absent in some models. In some cases $\lambda_i \, \sigma_i$ may represent a bare mass and, alternatively, some of the $\sigma_{NS}$, $\sigma_N$ and $\sigma_N$ singlet may correspond to the same singlet $\sigma$ or its conjugate. Similarly, the doublet $\chi$ may coincide with $H$ in some models. In general, the majoron will be a linear combination of the $CP$-odd parts of $H$, $\chi$ and $\sigma$, and then one can write
\begin{equation} \label{eq:majorongeneral}
J = \alpha_H H^a +\alpha_\chi \chi^a +\alpha_{\sigma} \sigma^a \equiv \alpha_H H^a +\alpha_\chi \chi^a +\alpha_{\sigma_{NS}} \sigma_{NS}^a + \alpha_{\sigma_N} \sigma_N^a + \alpha_{\sigma_S} \sigma_S^a  \, ,
\end{equation}
where the superscript $a$ refers to the $CP$-odd part of each scalar. The $\alpha_i$ coefficients encode the mixing in the $CP$-odd sector and can be easily computed in any given model. In some cases, it may occur that $\sigma_{NS} = \sigma_N$, $\sigma_S = \sigma_N^*$ or any other combination. This does not affect the majoron but does affect how our $\alpha_i$ coefficients must be taken. For example, if we have $\bar{N}^c S$ as a bare mass term,  the bilinear $\bar{N}^c N$ coupling to $\sigma$ and the bilinear $\bar{S}^c S$ coupling to $\sigma^*$, then we must take $\alpha_{\sigma_{NS}}=0$, $\alpha_{\sigma_{N}} = \alpha_\sigma$ and $\alpha_{\sigma_{S}} = -\alpha_\sigma$.
This fact can be expressed by the relation
\begin{equation}
    \alpha_\sigma = \frac{\lvert \alpha_{\sigma_{NS}} \rvert+\lvert \alpha_{\sigma_{N}}\rvert+ \lvert \alpha_{\sigma_{S}} \rvert}{3-\delta_{0 \,\alpha_{\sigma_{NS}}}-\delta_{0 \,\alpha_{\sigma_{N}}}-\delta_{0 \,\alpha_{\sigma_{S}}}} \, .
\end{equation}
After these preliminaries, the majoron coupling to neutral fermions in the gauge basis is given by 
\begin{align}
\label{eq:Abar}
\bar{A} = \frac{i}{2 \sqrt{2}} \begin{pmatrix}
0 & \alpha_H \, y_N  & \alpha_\chi\,  y_S \\
\alpha_H \, y_N^T & \alpha_{\sigma_N} \, \lambda_N  & \alpha_{\sigma_{NS}} \, \lambda \\
\alpha_\chi\,  y_S^T & \alpha_{\sigma_{NS}} \, \lambda^T &  \alpha_{\sigma_S}\,  \lambda_S 
\end{pmatrix} \equiv \frac{1}{2} \begin{pmatrix}
0 & \bar{A}_L \\
\bar{A}_L^T & \bar{A}_H
\end{pmatrix} \, ,
\end{align}
with 
\begin{align}
 \bar{A}_L = \frac{i}{ \sqrt{2}} \begin{pmatrix}
  \alpha_H \, y_N  & \alpha_\chi\,  y_S
 \end{pmatrix} \, , && \bar{A}_H = \frac{i}{ \sqrt{2}} \begin{pmatrix}
 \alpha_{\sigma_N} \, \lambda_N  & \alpha_{\sigma_{NS}} \, \lambda \\
 \alpha_{\sigma_{NS}}\, \lambda^T & \alpha_{\sigma_S}\,  \lambda_S 
\end{pmatrix} \, .
\end{align}
The charged scalar $\eta^+$ is also an admixture of two different fields,
\begin{equation}
\eta^+ = \beta_H \, H^{+} +\beta_\chi \, \chi^{+} \, ,
\end{equation}
where the $\beta_i$ coefficients parameterize the mixing. Then, the couplings of $\eta$ to a charged lepton and a neutral lepton are given, in the gauge basis, by
\begin{align}
\label{eq:Dbar}
  \bar{D}_R &= \frac{i}{ \sqrt{2}} \begin{pmatrix}
    \beta_H  \, y_N & \beta_\chi  \, y_S
  \end{pmatrix}
  \, ,
\end{align}
where, again, the absence of a coupling between $\eta$, the charged leptons and SM neutrinos has been used. Here, $\bar{D}_R$ is given in the $(N \, \, S)$ basis. With these expressions for the couplings, we have derived the general form of the majoron coupling to charged leptons in the Type-I Seesaw family, enabling us to adapt it for each specific model.

Let us now briefly discuss these results. We start with the matrices entering the $W$ and $Z$ contributions. Note that $\bar{A}$ is nothing but a Yukawa coupling, so $\bar{A}_{ij} \leq 1$. This implies that the dominant terms will be $\sim \frac{M_D^2}{M_F} \bar{A}_H$ and $\sim M_D \, \bar{A}_L$. For the first one, it is obvious that if we want to achieve a coupling that is not naturally suppressed by the light neutrino masses, we need $\bar{A}_H$ to alter the structure of $M_F^{-1}$. Regarding the second one, the reason is phenomenological. The majoron doublet admixture must be suppressed, as demanded by various experimental constraints. Of course, this suppression will be given by the VEV of $\sigma$. Then, the suppression will be of the order $v/v_\sigma$, $v_\chi/v_\sigma$, or some combination leading finally to, at least, the same suppression that we found in the first term. One can argue in the same way for the scalar diagrams, thus concluding that we need $\bar{A}$ to alter the structure of $M_F^{-1}$ if we want sizable rates for flavor processes involving the majoron.

\bibliographystyle{utphys}
\providecommand{\href}[2]{#2}\begingroup\raggedright\endgroup

\end{document}